\documentclass[preprint,12pt]{elsarticle}

\newcommand{\abs}[1]{\ensuremath{\left\vert#1\right\vert}}

\def\D{\mathrm{d}}

\usepackage{amssymb}
\usepackage{amsmath}
\usepackage{amsfonts}
\usepackage{booktabs}
\usepackage{enumitem}
\usepackage{subfigure} 
\usepackage{multirow}
\usepackage{diagbox}
\usepackage{float}
\usepackage{bm}
\usepackage{relsize}
\usepackage{color}
\usepackage{lmodern}
\usepackage{trfsigns}
\usepackage{nomencl} 
\usepackage{framed} 
\usepackage[colorlinks=true]{hyperref}
\makenomenclature

\usepackage{lineno}

\journal{ }
\begin{document}

\begin{frontmatter}

\nomenclature[ag, 01]{$\alpha$}{angle of rotation}
\nomenclature[a, 01]{$A$}{participation factors}
\nomenclature[bg, 01]{$\beta$}{kurtosis}
\nomenclature[c, 01]{$c_n$}{cumulant of $n^{\text{th}}$ order}
\nomenclature[d, 02]{$\bm{D}$}{damping matrix}
\nomenclature[d, 03]{$\bm{d}$}{diagonalized damping matrix}
\nomenclature[d, 05]{$\zeta$}{damping coefficient}
\nomenclature[e, 01]{$\bm{E}$}{elasticity matrix}
\nomenclature[ep, 01]{$\varepsilon$}{strain}
\nomenclature[f, 01]{$f$}{frequency}
\nomenclature[f, 02]{$f_n$}{frequency arguments}
\nomenclature[f, 03]{$f_s$}{sampling frequency}
\nomenclature[g, 01]{$G_2^{(x)}(f)$}{one-sided PSD of $x$}
\nomenclature[gg, 01]{$\gamma$}{skewness}
\nomenclature[h, 02]{$H^{(xy)}(f)$}{frequency-response function from $x$ to $y$}
\nomenclature[h, 02]{$H^{(\bm{xq})}(f)$}{general FRF from multivariate input $\bm{x}$ to modal solution $\bm{q}$}
\nomenclature[i, 01]{$i$}{complex number}
\nomenclature[k, 02]{$\bm{K}$}{stiffness matrix}
\nomenclature[k, 03]{$\bm{k}$}{diagonalized stiffness matrix}
\nomenclature[n, 01]{$n$}{order}
\nomenclature[n, 02]{$N$}{number of degrees of freedom (DOFs)}
\nomenclature[n, 03]{$N_r$}{number of modes for response analysis}
\nomenclature[n, 04]{$N_x$}{number of inputs}
\nomenclature[n, 05]{$N_{\sigma}$}{number of stresses}
\nomenclature[m, 01]{$m_n$}{statistical moment of $n^{\text{th}}$ order}
\nomenclature[m, 02]{$\bm{M}$}{mass matrix}
\nomenclature[m, 03]{$\bm{m}$}{diagonalized mass matrix}
\nomenclature[mg, 01]{$\mu$}{mean}
\nomenclature[mg, 02]{$\bm{\mu}$}{mean vector}
\nomenclature[mg, 03]{$\mu_n$}{central moment of $n^{\text{th}}$ order}
\nomenclature[mg, 04]{$\mu_n^{(\bm{x})}$}{$n^{\text{th}}$-order central moment for multivariate input $\bm{x}$}
\nomenclature[mg, 05]{$\hat{\mu}_4^{(\bm{x})}$}{projected fourth-order central moment for multivariate input $\bm{x}$}
\nomenclature[o, 01]{$\omega$}{angular frequency}
\nomenclature[o, 02]{$\omega_0$}{eigenfrequencies}
\nomenclature[p, 02]{$p(x)$}{PDF for variable $x$}
\nomenclature[p, 03]{$p_g(x)$}{Gaussian PDF for variable $x$}
\nomenclature[ph, 01]{$\phi$}{mode shape / eigenvector}
\nomenclature[ph, 02]{$\Phi$}{modal matrix}
\nomenclature[q, 02]{$q(t)$}{modal coordinates / modal solution}
\nomenclature[r, 01]{$r$}{index for modes}
\nomenclature[r, 02]{$r_2^{(x)}(\tau)$}{auto-correlation function of $x$}
\nomenclature[s, 01]{$S_n(f_1,...,f_{n-1})$}{spectrum of $n^{\text{th}}$ order}
\nomenclature[s, 02]{$s(t)$}{nodal displacement vector}
\nomenclature[s, 03]{$\bm{S}$}{differential operator matrix}
\nomenclature[sg, 01]{$\sigma$}{standard deviation}
\nomenclature[sg, 02]{$\sigma^2$}{variance}
\nomenclature[sg, 03]{$\sigma_{x_{u_1}x_{u_2}}^2$}{covariance between $x_{u_1}$ and $x_{u_2}$}
\nomenclature[sg, 04]{$\bm{\Sigma}$}{covariance matrix}
\nomenclature[sg, 05]{$\bm{\tilde{\Sigma}}$}{covariance matrix in Voigt form}
\nomenclature[sg, 06]{$\bm{\sigma}$}{stress tensor}
\nomenclature[t, 01]{$t$}{time}
\nomenclature[u, 01]{$u$}{index for univariate processes}
\nomenclature[u, 02]{$U$}{number of processes}
\nomenclature[u, 01]{$v$}{index for response processes}
\nomenclature[u, 02]{$V$}{number of response processes}
\nomenclature[x, 01]{$X(t)$}{stochastic process}
\nomenclature[x, 04]{$x(t)$}{random excitation (realization)}
\nomenclature[x, 05]{$X(f)$}{Fourier transform of $x$}
\nomenclature[y, 01]{$y(t)$}{random response (realization)}

\nomenclature[21, 01]{$E[\cdot]$}{expected value}
\nomenclature[22, 02]{$\abs{\cdot}$}{absolute value}
\nomenclature[23, 03]{$[\cdot]^*$}{complex conjugated}
\nomenclature[24, 04]{$[\cdot]^T$}{transpose}
\nomenclature[24, 05]{$\otimes$}{tensor product}
\nomenclature[24, 06]{$\oslash$}{elementwise division}
\nomenclature[24, 07]{$^{\otimes}$}{tensor power}
\nomenclature[24, 08]{$\Re[\cdot]$}{real part} 
\nomenclature[24, 09]{$\mathcal{F}\{\cdot\}$}{Fourier transform}
\nomenclature[24, 09]{$\mathcal{F}^{-1}\{\cdot\}$}{inverse Fourier transform}
\nomenclature[101, 01]{DOF}{degrees-of-freedom}
\nomenclature[102, 01]{FE}{finite elemente}
\nomenclature[103, 01]{FRF}{frequency response function}
\nomenclature[104, 01]{HOS}{higher-order spectra}
\nomenclature[110, 01]{MIMO}{multiple input multiple output}
\nomenclature[112, 01]{NSM}{non-stationarity matrix}
\nomenclature[118, 01]{PDF}{probability density function}
\nomenclature[120, 01]{PSD}{power spectral density}

\author{Arvid Trapp \corref{cor1}}
\ead{arvid.trapp@hm.edu}
\author{Peter Wolfsteiner}
\address{University of Applied Sciences Munich,\\ Department of Mechanical, Automotive and Aeronautical Engineering, \\ Dachauer Strasse 98b, 80335 Munich, Germany\fnref{label3}}

\title{Fast assessment of non-Gaussian inputs in structural dynamics exploiting modal solutions}

\begin{abstract}
In various technical applications, assessing the impact of non-Gaussian processes on responses of dynamic systems is crucial. While simulating time-domain realizations offers an efficient solution for linear dynamic systems, this method proves time-consuming for finite element (FE) models, which may contain thousands to millions of degrees-of-freedom (DOF). Given the central role of kurtosis in describing non-Gaussianity --- owing to its concise, parametric-free and easily interpretable nature --- this paper introduces a highly efficient approach for deriving response kurtosis and other related statistical descriptions. This approach makes use of the modal solution of dynamic systems, which allows to reduce DOFs and responses analysis to a minimum number in the modal domain. This computational advantage enables fast assessments of non-Gaussian effects for entire FE models. Our approach is illustrated using a simple FE model that has found regular use in the field of random vibration fatigue.
\end{abstract}


\end{frontmatter}

\printnomenclature

\section{Introduction}
Structural dynamics occupies a central role across various technical fields and primarily deals with understanding how structures behave under dynamic loading. One of its core applications is solving classic input-system-output problems, e.g.~for explaining structural behavior and responses. Solving these problems is not only critical for the design and analysis of structures but also for predicting their long-term performance and safety under varying loads. 

One of the most relevant methods employed in structural dynamics is the finite element (FE) method. FE models are able to represent complex physical systems, allowing for detailed analysis of designs using cost-efficient numerical models. However, the inherent discretization of these models often comprises thousands to millions of degrees of freedom (DOFs), among others, necessary for accurately reflecting stress concentrations. In the context of input-system-output schemes, this level of detail bears significant computational challenges, especially when lengthy dynamic loads and entire FE models, i.e.~an extensive number of outputs, need to be solved and evaluated. In addressing this challenge, modal analysis --- respectively the modal perspective --- has emerged as a powerful approach. By focusing on the relevant modes of vibration and their contributions to the overall system behavior, this method offers a way to significantly reduce computational effort while maintaining accuracy of the models. 

The field of vibration fatigue has contributed a set of valuable methodologies to structural dynamics. Being a relatively old engineering discipline, fatigue refers to the continuous degradation of materials due to cyclic loading. In vibration fatigue, cyclic loading of materials is decisively driven by structural resonances. As such, fatigue assessments are conducted to prevent structural failures due to fatigue, incorporating structural dynamics at its core. Such an assessment can be carried out following a sampling-based 'time-domain' and a statistical-based 'frequency-domain' approach. In a sampling-based approach time-domain realizations are processed for linear systems, e.g.~using the frequency response function (FRF) matrices obtainable from FE models. The resulting response time series may then be processed by cycle counting and for multiaxial fatigue criteria to predict structural lifetimes. Because of the many DOFs of FE models and the computationally expensive post-processing, pursuing a sampling-based approach requires substantial amounts of time. This often hinders making full use of available load data (e.g.~recorded data on existing and prototyping structures), but also the process of optimizing and validating designs. A statistical-based assessment is substantially faster utilizing a set of clever approaches. These include the use of spectral densities, i.e.~statistical characterizations of dynamic loads and structural responses. The most commonly employed spectral density is the power spectral density (PSD), which substantially lowers the amount of data to be processed in linear systems theory. Further, on the basis of PSD-based characterizations a variety of analytical and empirical models have been introduced which estimate peak amplitude distributions for narrowband processes \cite{Lutes.2004}, expected peak amplitudes \cite{Cartwright.1956,Davenport.1964}, and rainflow-counting collectives for broadband processes \cite{Dirlik.1985,Benasciutti.2004}. Also a series of multiaxial fatigue criteria have been adjusted to run on PSD-based statistical characterizations \cite{Nieslony.2007,Benasciutti.2015,Mrsnik.2016,Carpinteri.2017}. These methodologies provide the foundation for a statistical-based approach to a fatigue assessment --- closely linking statistical characterizations with structural dynamics, providing significantly faster and in principle more robust results. Finally to mention, \cite{Braccesi.2016, Braccesi.2017} have shown, that these methods can be efficiently implemented in the analysis of FE models with almost negligible computational effort by transferring the statistical characterization to the modal domain. However, the flip side of the statistical-based approach is connected to its PSD-based characterization. The PSD matrix provides a full statistical characterization exclusively for stationary Gaussian processes. When we validate this assumption on the basis of recorded vibration loads, in most applications we find complex dynamic loading that significantly deviates from stationarity and Gaussianity. In such cases a statistical-based 'frequency-domain' approach tends to produce significantly non-conservative results \cite{Wolfsteiner.2013,Palmieri.2017,Decker.2018b}, which risks inappropriate structural designs. This often undermines the practical use of a statistical-based 'frequency-domain' approach. 

Stationary Gaussian processes are the result of the central limit theorem of statistics and fundamental to many physical processes subjected to randomness. In human-made systems, however, for various reasons we often find deviations from stationarity and Gaussianity, which mostly relates to varying and mixed load conditions. For example, if we measure loads on a train, these will depend on a wide variety of effects. To name a few, on environmental conditions, the traction drive concept, the human controlling the train, the speed of the train, the tracks the train rides, the trains payload, the components that are attached to the train and many more factors. From a signal processing view, this will affect dynamic loads by phenomena like amplitude-modulation, frequency-modulation and superimposed deterministic loading, conflicting the demands of a statistical-based fatigue assessment. For such time-dependent processes, where the intensity and its frequency decomposition varies, the PSD represents the average intensity for frequency and conceals any information about its time evolution. As such, it insufficiently characterizes processes that have a non-stationary evolution or are subjected to instantaneous events. 
To better account for such load conditions while maintain the advantages of a statistical approach, this paper demonstrates how higher-order statistical characterizations, such as the popular kurtosis, can be efficiently obtained by a modal approach for entire FE models. Kurtosis concisely quantifies how a probability density function (PDF) differs from a stationary Gaussian process of same standard deviation.
Although kurtosis has been widely adopted in numerous proposals to capture the effects of non-Gaussianity in vibration fatigue \cite{Winterstein.1988,Benasciutti.2006b, Cianetti.2018b}, few \cite{Kihm.2013,Trapp.2021b,Cui.2022} have discussed the practical derivation of the requisite response kurtoses. I.e.~how kurtosis 'transfers' through structural dynamics. The herein proposed approach fills this gap by enabling the calculation of responses kurtosis with minimal computational effort and as such allowing to efficiently evaluate the impact of non-Gaussian inputs in structural dynamics.  

This paper begins by establishing the foundational concepts of statistical characterizations of random processes and their application in structural dynamics (Sec.~\ref{sec:fundamentals_stochasticAssessmentAndStructuralDynamics}). The main content (Sec. \ref{sec:Main_MA4Kurtosis}) then progresses to extend the modal approach to fourth-order statistical characterizations, focusing on the multivariate fourth-order moment, alongside using the modal solutions for statistical response analysis. Subsequently the new content is applied in Sec.~\ref{sec:Validation}, offering visual demonstrations of the proposed methodologies' effectiveness. The paper concludes by summarizing the findings, emphasizing the advantages of utilizing modal solutions in the analysis of non-Gaussian inputs on linear systems, and suggesting directions for future research.

\section{Stochastic assessment of random processes and its processing in structural dynamics}
\label{sec:fundamentals_stochasticAssessmentAndStructuralDynamics}
This section delves into the basic concepts and methodologies required for the main content of this paper, all of which belong to the stochastic assessment of random processes and its processing for structural dynamics. Sec.~\ref{subsec:F_univariate} begins with the univariate stochastic assessment of random variables introducing concepts like the kurtosis. Building on this, the well-established second-order multivariate stochastic assessment (Sec.~\ref{subsec:F_multivariate}) is introduced, extending the principles of the univariate assessment to multiple random variables. Multivariate analysis is the precondition for correctly  characterizing the modal solution of realistic, multi-modal structures. Lastly, modal analysis is introduced in the final Sec.~\ref{subsec:F_StrucDyn} of this introductory passage. 
\label{sec:statisticalAssessment}
\subsection{Univariate stochastic assessment}
\label{subsec:F_univariate}
Univariate stationary vibration conforming the central limit theorem follows the Gaussian probability density function (PDF), 
\begin{equation} \label{eq:F_UV_PDensityGauss}
p_g(x) = \dfrac{1}{\sqrt{2 \pi \sigma^2}} \; \mathrm{e}^{\left(-\dfrac{(x-\mu)^2}{2 \, \sigma^2}\right)}
\end{equation}
where $x$ generally represents a variable that is subjected to random influences. The Gaussian PDF is fully characterized by mean $\mu$ and variance $\sigma^2$. Both belong to a wider range of statistical descriptors termed moments $m_n$, measuring spread about zero, and central moments $\mu_n$ measuring spread about the mean $\mu$. In particular, the variance is the second-order central moment (Eq.~\ref{eq:F_UV_NOrderCentMoments}) $\sigma^2=\mu_2$, while the mean is the first statistical moment $\mu=m_1$ (Eq.~\ref{eq:F_UV_NOrderStatMoments}).
\begin{subequations} \label{eq:F_UV_NOrderMoments}
\begin{equation} \label{eq:F_UV_NOrderStatMoments}
m_{n}^{(x)}=E[X^{n}(t)] = \int _{-\infty }^{\infty }x^{n} \, p(x) \; dx \\
\end{equation}
\begin{equation} \label{eq:F_UV_NOrderCentMoments}
\mu_{n}^{(x)}=E\big[(X(t)-E[X(t)])^{n}\big] =\int_{-\infty }^{\infty }(x-{\mu})^{n} \, p(x) \; dx
\end{equation}
\end{subequations}
To describe if and to which degree a PDF $p(x)$ deviates from a stationary Gaussian process with PDF $p_g(x)$, popular descriptors are the skewness $\gamma$ and the kurtosis $\beta$. These are standardized higher-order moments (Eq.~\ref{eq:F_UV_NOrderCentMoments}) characterizing a PDF $p(x)$ independently of the standard deviation $\sigma = \sqrt{\mu_2}$
\begin{subequations} \label{eq:F_UV_GammUndBeta}
\begin{equation} \label{eq:F_UV_Gamma}
\gamma =\frac{\mu _{3}}{\sigma ^{3}}=\frac{\mu _{3}}{(\mu _{2})^{\frac{3}{2}}} 
\end{equation}
\begin{equation} \label{eq:F_UV_Beta}
 \beta =\frac{\mu _{4}}{\sigma ^{4}}=\frac{\mu _{4}}{(\mu _{2})^{2}}
\end{equation}
\end{subequations}
Skewness $\gamma $ (third order) quantifies the asymmetry of a PDF, whereas kurtosis $\beta$ (fourth order) measures the width of spread --- capable of indicating rare extreme events. Especially this latter characteristic is particularly critical for many applications. For instance, in the context of vibration fatigue, the lifetime of structures is inversely related to stress amplitudes, diminishing according to a power law (stress-life curves). 
Cumulants $c_n$ are an alternative to moments that provide beneficial mathematical properties and a favorable representation for analyzing statistical dependence. Here, their main interest lies in providing the moment contribution that deviates from a stationary Gaussian process --- in contrast to Eqs.~\eqref{eq:F_UV_GammUndBeta}, cumulants provide non-normalized statistical characterizations of non-Gaussianity. The cumulants' concept requires to estimate them in terms of central moments, whose relation is given by Bell polynomials \cite{Trapp.2023f}. Second- to fourth-order cumulants are defined for central moments by
\begin{align} \label{eq:F_UV_Cumulants}
\begin{split}
c_2 &= \mu_{2} \\
c_3 &= \mu_{3} \\
c_4 &= \mu_{4} - 3 \mu_2^2 \\
\end{split} 
\end{align}
Higher-order $n>2$ moments and cumulants belong to higher-order statistics. One of its concepts is spectral analysis $S_n^{(x)}(f_1,...,f_{n-1})$, which provides the frequency-domain decomposition of central moments 
\begin{equation}
\label{eq:F_UV_HOParsvl}
\mu_n^{(x)} = \int_{-\infty}^{\infty}...\int_{-\infty}^{\infty} S_n^{(x)} (f_1,...,f_{n-1}) \, \D f_1... \D f_{n-1}
\end{equation}
These spectral decomposition are required when statistical characterizations are processed for linear systems --- to estimate structural responses in a dynamic analysis (Sec.~\ref{subsec:F_StrucDyn}). 
$S_n^{(x)} (f_1,...,f_{n-1})$ of orders $n>2$ are termed higher-order spectra (HOS) \cite{Nikias.1993b}. However, the most relevant spectrum is certainly the second-order power spectral density (PSD),
\begin{subequations} \label{eq:F_UV_PSD}
\begin{align}
\text{via auto-correlation function } r_2^{(x)}(\tau): & \hspace{10mm} S_{2}^{(x)}(f) = \int_{-\infty}^{\infty} r_2^{(x)}(\tau) \, \mathrm{e}^{-i2\pi f \tau} \, \D \tau \\
\text{via Fourier transform } X(f): & \hspace{10mm} S_{2}^{(x)}(f) = \lim_{T\to \infty} \frac{1}{T} E[X(f)X^{*}(f)]
\end{align}
\end{subequations}

Because of its inherent symmetry, commonly one-sided PSDs are used, 
\begin{equation} \label{eq:F_UV_PSDoneSided}
G_2^{(x)}(f)= \left\{ \begin{array}{cc}
2 S_2^{(x)}(f)  & \text{for } f>0\\
S_2^{(x)}(f) & \text{for } f=0 \\
0 & \text{for } f <0 
\end{array} \right.
\end{equation} 
The PSD spectrally decomposes the variance or formally the second-order central moment 
\begin{equation} \label{eq:F_UV_PSDintegration}
\sigma_x^2 = \mu_2^{(x)} = \int_{-\infty}^{\infty}S_2^{(x)}(f) \, \D f = \int_{0}^{\infty}G_2^{(x)}(f) \, \D f
\end{equation}
The HOS related to the kurtosis $\beta^{(x)}$ (Eq.~\ref{eq:F_UV_Beta}) is the fourth-order trispectrum $S_{4}^{(x)}(f_1,f_2,f_3)$, combining Eqs.~\eqref{eq:F_UV_Beta} and \eqref{eq:F_UV_HOParsvl} resp.~Eq.~\eqref{eq:F_UV_PSDintegration} for the denominator
\begin{equation}
\label{eq:F_UV_Kurtosis}
\beta^{(x)} = \frac{\mu_4^{(x)}}{(\mu_2^{(x)})^2} = \frac{\int_{-\infty}^{\infty}\int_{-\infty}^{\infty}\int_{-\infty}^{\infty} S_{4}^{(x)}(f_1,f_2,f_3) \, \D f_1 \D f_2 \D f_3}{(\int_{-\infty}^{\infty}S_2^{(x)}(f) \,\D f)^2} 
\end{equation}
Additionally, \cite{Trapp.2021b} proposes the non-stationarity matrix (NSM), providing an alternative spectral representation of the fourth-order moment $\mu_4$ and cumulant $c_4$, but also the kurtosis $\beta$. The NSM provides easier representation and interpretability for the fourth-order characterizations than the trispectrum, as it has only two-frequency arguments, analyzing the time-correlation between the two frequency components. 
\subsection{Multivariate stochastic assessment}
\label{subsec:F_multivariate}
Extending the analysis to multivariate processes, the number of process variables $\bm{x} = \{x_1,...,x_U\}$ shall be defined by $U$, so that $u = 1,...,U$ assigns the underlying univariate processeses. The multivariate Gaussian PDF $p_g(\bm{x})$ is analogously to Eq.~\eqref{eq:F_UV_PDensityGauss} fully characterized by moments of first and second order $n = \{1,2\}$ --- therefore mean and variance are extended to mean vector $\mu^{(\bm{x})}$ and covariance matrix $\bm{\Sigma}^{(\bm{x})}$. It is important to note that all scalar-value descriptions extended to a multivariate representation retain their non-bold letters, while the now multi-dimensional nature is indicated by the bold referencing variable in parenthesis.
\begin{equation} \label{eq:F_MVM_PDensityGauss}
p_g(\bm{x}) = p_g(x_1,...,x_U) = \dfrac{1}{\sqrt{(2 \pi)^U \abs{\bm{\Sigma}^{(\bm{x})}}}} \; \mathrm{e}^{\left(-\dfrac{(\bm{x}-\mu^{(\bm{x})})^T\bm{\Sigma}^{(\bm{x})^{-1}}(\bm{x}-\mu^{(\bm{x})})}{2}\right)}
\end{equation}
Covariance describes the average linear correlation,
\begin{equation} \label{eq:F_MVM_NOrderMoments}
\mu_{2}^{(x_{u_1,u_2})}= \sigma_{x_{u_1,u_2}} = E\big[(X_{u_1}(t)-E[X_{u_1}(t)])(X_{u_2}(t)-E[X_{u_2}(t)]) \big]
\end{equation}
which in case of $u_1 = u_2$, i.e.~the diagonal of the covariance matrix, gives the prior introduced univariate second-order central moments (Eq.~\ref{eq:F_UV_NOrderCentMoments})
\begin{equation} \label{eq:F_MVM_NOrderMoments_Variance}
\text{for } u_1 = u_2: \;\;\;\;\;\; \mu_{2}^{(x_{u})} = \sigma^2_{x_{u}} \; \hat{=} \; \mu_{2}^{(x_{u,u})} = E\big[(X_u(t)-E[X_u(t)])^2\big]
\end{equation}
This second-order characterization is assembled in the symmetric covariance matrix ($U\text{ x }U$),
\begin{equation} \label{eq:F_MVM_CovMatrix}
\bm{\Sigma}^{(\bm{x})} = 
\begin{bmatrix}
\sigma^2_{x_1} & \sigma_{x_{1,2}} & \cdots & \sigma_{x_{1,U}} \\
\sigma_{x_{2,1}} & \sigma^2_{x_2} & \cdots & \sigma_{x_{2,U}} \\
\vdots & \vdots & \ddots & \vdots  \\
\sigma_{x_{U,1}} & \sigma_{x_{U,2}} & \cdots & \sigma^2_{x_U} \\
\end{bmatrix} 
= \mu_2^{(\bm{x})} = 
\begin{bmatrix}
\mu_{2}^{(x_1)} & \mu_{2}^{(x_{1,2})} & \cdots & \mu_{2}^{(x_{1,U})} \\
\mu_{2}^{(x_{2,1})} & \mu_{2}^{(x_2)} & \cdots & \mu_{2}^{(x_{2,U})} \\
\vdots & \vdots & \ddots & \vdots  \\
\mu_{2}^{(x_{U,1})} & \mu_{2}^{(x_{U,2})}  & \cdots & \mu_{2}^{(x_U)} \\
\end{bmatrix} 
\end{equation}
whose spectral decomposition is provided by the PSD matrix
\begin{equation} \label{eq:F_MVM_PSDMatrix}
G_2^{(\bm{x})}(f) = 
\begin{bmatrix}
G_2^{(x_1)}(f) & G_2^{(x_{1,2})}(f) & \cdots & G_2^{(x_{1,U})}(f)  \\
G_2^{(x_{2,1})}(f) & G_2^{(x_2)}(f) & \cdots & G_2^{(x_{2,U})}(f)  \\
\vdots & \vdots & \ddots & \vdots \\
G_2^{(x_{U,1})}(f)  & G_2^{(x_{U,2})}(f) & \cdots & G_2^{(x_U)}(f) 
\end{bmatrix}
\end{equation}
In accordance with the covariance matrix, the PSD matrix assembles on its diagonal the univariate PSDs (Eq.~\ref{eq:F_UV_PSDintegration}) and on its off-diagonals cross spectral densities (CPSD), comp.~Eq.~\eqref{eq:F_UV_PSD}
\begin{subequations} \label{eq:F_MVM_CSDuPeriodogramm}
\begin{align}
S_2^{(x_{u_1,u_2})}(f) &= \int_{-\infty}^{\infty} r_2^{(x_{u_1,u_2})}(\tau) \, \mathrm{e}^{-i2\pi f \tau} \D\tau; \\
S_2^{(x_{u_1,u_2})}(f) &= \lim_{T\to \infty} \frac{1}{T} E[X_{u_1}(f)X_{u_2}^{*}(f)] 
\end{align}
\end{subequations}
The PSD matrix is Hermitian symmetric, i.e.~$S_2^{(x_{u_1,u_2})}(f)$ is the complex-conjugated of $S_2^{(x_{u_2,u_1})}(f)$. They relate to the covariances by their integral of the real part.
Consequently, the spectral pendant to $\bm{\Sigma}^{(\bm{x})}$ (Eq.~\ref{eq:F_MVM_CovMatrix}) is given by 
\begin{align} \label{eq:F_MVM_Mom2IntPSDMatrix}
\begin{split}
\mu_2^{(\bm{x})} = \bm{\Sigma}^{(\bm{x})} =& 
\mathop{\mathlarger{\int}}_0^{\infty}  G_2^{(\bm{x})} \D f \\
\mu_2^{(\bm{x})} = 
\begin{bmatrix}
\mu_{2}^{(x_1)} & \mu_{2}^{(x_{1,2})} & \cdots & \mu_{2}^{(x_{1,U})} \\
\mu_{2}^{(x_{2,1})} & \mu_{2}^{(x_2)} & \cdots & \mu_{2}^{(x_{2,U})} \\
\vdots & \vdots & \ddots & \vdots  \\
\mu_{2}^{(x_{U,1})} & \mu_{2}^{(x_{U,2})}  & \cdots & \mu_{2}^{(x_U)} \\
\end{bmatrix}  =&  
\begin{cases} \mathop{\mathlarger{\int}}_0^{\infty} G_2^{(x_{u})}(f) \D f & \forall u = u_1 = u_2 \\ \\ \mathop{\mathlarger{\int}}_0^{\infty} \Re\big[G_2^{(x_{u_1,u_2})}(f)\big] \D f & \forall u_1 \neq u_2 \\ \end{cases} 
\end{split}
\end{align}
whereby this equation conditions element-wise integration of the real part. \\
To end this section, note that the covariance matrix $\bm{\Sigma}^{(\bm{x})} = \mu_2^{(\bm{x})}$ can also be defined as tensor of rank two
\begin{equation} \label{eq:F_MVM_CovMatrix_Tensor}
\begin{split}
\mu_2^{(\bm{x})} = \bm{\Sigma}^{(\bm{x})} = E\big[(\bm{X}(t)-E[\bm{X}(t)])^{2}\big] = E\big[(\bm{x}-\mu^{(\bm{x})})^{\otimes 2}\big] \\
= E\big[(\bm{x}-\mu^{(\bm{x})}) \otimes (\bm{x}-\mu^{(\bm{x})})\big]
\end{split}
\end{equation}
This becomes important when extending the analysis to higher orders $n > 2$ in the main section (Sec.~\ref{sec:Main_MA4Kurtosis}), e.g.~for the multivariate fourth-order moment and its related kurtosis. 
\subsection{Structural dynamics and modal approach}
\label{subsec:F_StrucDyn}
\begin{figure}[h]
\centering
\includegraphics[keepaspectratio,width=\textwidth,height=\textheight]{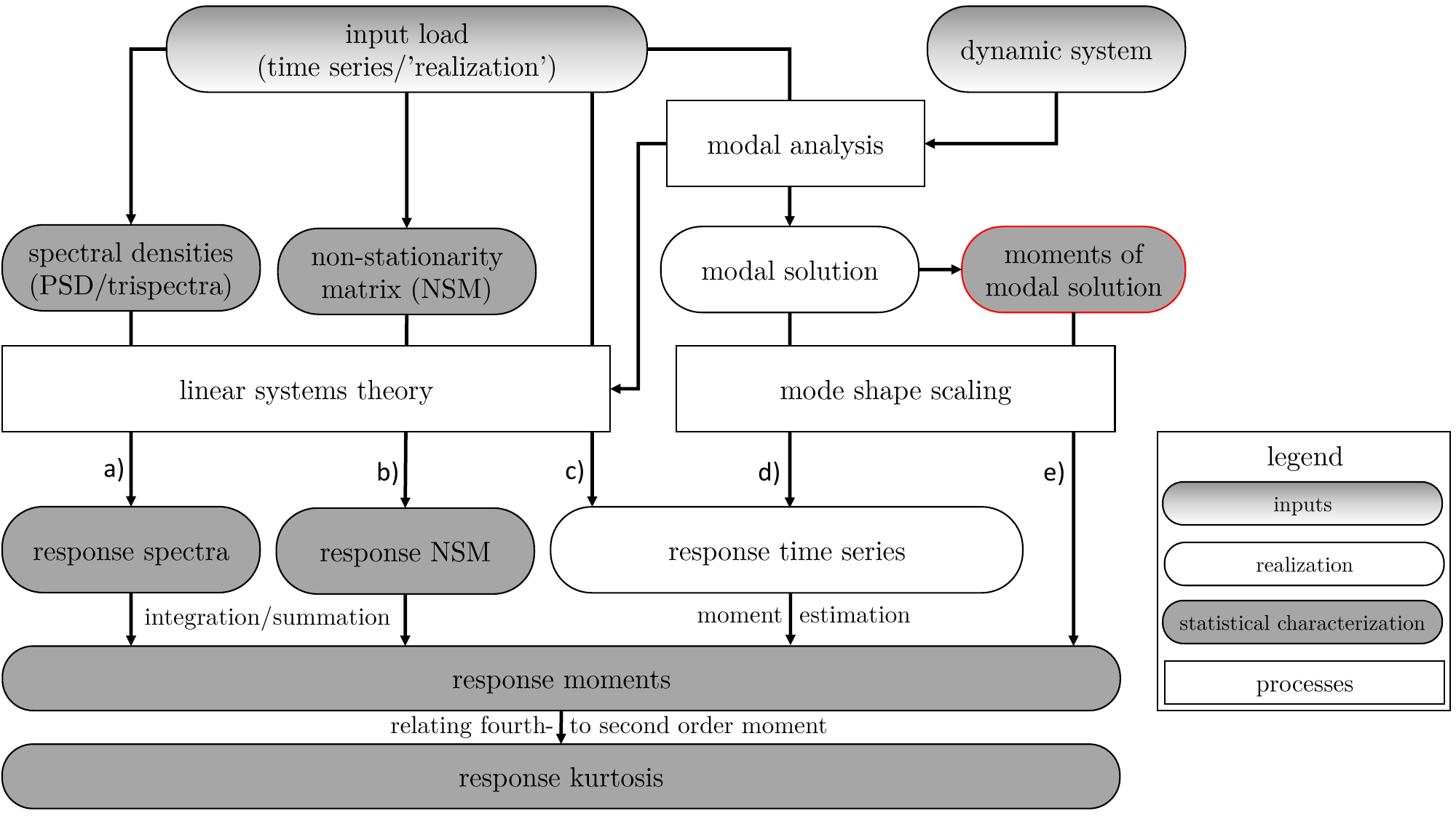}
 \caption{Schematic overview of different paths to obtain response kurtoses: Using linear systems theory, via a) spectral densities including the trispectrum, b) the non-stationarity matrix (NSM), and c) response time series and subsequent moment estimation, alternatively using mode shape scaling, via d) response time series and e) second- and fourth-order moments of the modal solution, presented within this paper}
\label{fig:Approaches4DerivingReponseKurtoses}
\end{figure}

In structural dynamics linear systems theory plays a crucial role, establishing the relation between a set of $U$ input variables $\bm{x}$ to series of $V$ output variables $\bm{y}$ for linear time-invariant systems by the FRF matrix $H^{(\bm{xy})}(f)$. In case of the second-order PSD matrix responses are obtained by $G_2^{(\bm{y})}(f) = H^{(\bm{xy})^{*^{T}}}(f) G_2^{(\bm{x})}(f) H^{(\bm{xy})}(f)$, represented in matrix-form by 
\begin{equation} \label{eq:F_MVM_LinSysTheo2Ordnung}
\begin{split}
\left[ \begin{array}{rrr}
G_2^{(y_{1})}(f) & \cdots & G_2^{(y_{1,V})}(f) \\
\vdots & \ddots & \vdots \\
G_2^{(y_{V,1})}(f) & \cdots & G_2^{(y_V)}(f) \\
\end{array} \right] = .. \\
.. = \left[ \begin{array}{rrr}
H_{x_1y_1}^*(f) & \cdots & H_{x_Uy_1}^*(f) \\
\vdots & \ddots & \vdots \\
H_{x_1y_V}^*(f) & \cdots & H_{x_Uy_V}^*(f) \\
\end{array}\right] 
\left[ \begin{array}{rrr}
G_2^{(x_1)}(f) & \cdots & G_2^{(x_{1,U})}(f) \\
\vdots & \ddots & \vdots  \\
G_2^{(x_{U,1})}(f) & \cdots & G_2^{(x_U)}(f) \\
\end{array}\right] 
\left[ \begin{array}{rrr}
H_{x_1y_1}(f) & \cdots & H_{x_1y_V}(f) \\
\vdots & \ddots & \vdots \\
H_{x_Uy_1}(f) & \cdots & H_{x_Uy_V}(f) \\
\end{array}\right]  
\end{split} 
\end{equation}  
Modal analysis provides a very efficient approach to derive FRFs as required for Eq.~\eqref{eq:F_MVM_LinSysTheo2Ordnung}. And more broadly speaking, for understanding the vibrational behavior of mechanical systems, especially within the context of FE models. 
Modal analysis is based on the linear equation of motion (EOM)
\begin{align} \label{eq:EqOfMot}
	\bm{M} \,\ddot{\bm{s}}(t) +  \bm{D} \,\dot{\bm{s}}(t) + \bm{K} \,\bm{s}(t) = \bm{A}\, \bm{x}(t)
\end{align}
where $\bm{K}$, $\bm{D}$ and $\bm{M}$ represent the structural stiffness, damping and mass matrix (each dimensioned by $N$) and  the nodal displacement vector $\bm{s}(t)$. Further, $\bm{x}(t)$ represents a time-dependent force or motion excitation vector (dimension $U$) and matrix $\bm{A}$ its participation factors. At the core of modal analysis lies the eigenvalue problem of dimension $N$, which is described by
\begin{equation}\label{eq:F_SD_EigenvalueProblem}
	\left(\bm{K} - {\omega_r}^2 \bm{M}\right)\;\bm{\phi_r} = \bm{0}
\end{equation}
Its solution yields a series of eigenvalues $\{{\omega_1}, ..., {\omega_r}, ..., {\omega_N}\}$ and eigenvectors $\{\bm{\phi_1}, ..., \bm{\phi_r}, ..., \bm{\phi_N}\}$, representing a combination of natural frequencies and corresponding modal displacement shapes of the vibration system, neglecting the influence of damping. These mode shapes are essential for understanding the structures' properties. Under the assumption of modal damping they can be used to analyze its vibrational behavior. Typically, the number of mode shapes $N_{r}$ considered in such analyses is much less than the total number of nodal DOFs $N$ in the corresponding detailed FE models. This reduction from $N$ to $N_{r}$ simplifies the analysis and speeds up computations without significantly sacrificing accuracy. The system's response can be effectively assessed by turning from nodal $\bm{s}(t)$ to modal coordinates $\bm{q}(t)$. This requires the modal matrix $\bm{\Phi} = [\bm{\phi_1}, ..., \bm{\phi_r} ...,\bm{\phi_N}]$, which organizes the modal displacement shapes with: $\bm{s}(t) = \bm{\Phi} \bm{q}(t)$. When assuming modal damping, the equation of motion \eqref{eq:EqOfMot} can be uncoupled to:
\begin{align} \label{eq:EqOfMot_modal}
	\underbrace{\left( \bm{\Phi}^T \bm{M} \bm{\Phi} \right)}_{\bm{m}} \ddot{\bm{q}}(t) +  \underbrace{\left( \bm{\Phi}^T \bm{D} \bm{\Phi} \right)}_{\bm{d}} \dot{\bm{q}}(t) + \underbrace{\left( \bm{\Phi}^T \bm{K} \bm{\Phi} \right)}_{\bm{k}} \bm{q}(t) = \underbrace{\left(\bm{\Phi}^T \bm{A} \right)}_{\bm{\Phi}^{(x)}} \bm{x}(t)
\end{align}
with the diagonalized matrices $\bm{m}$, $\bm{d}$ and $\bm{k}$ and the modal participation matrix $\bm{\Phi}^{(x)}$. This gives a representation of the EOM by decoupled single-degree-of-freedom (SDOF) systems. Solving this for forced vibration in the frequency domain:
\begin{align} \label{eq:F_SD_EqOfMotModal}
	\left(-\omega^2 \bm{m} + i\omega \bm{d} + \bm{k}\right) \bm{Q}(\omega) = \bm{\Phi}^{(x)} \bm{X}(\omega)
\end{align}
yields the FRF matrix $H^{\bm{(xq)}}(\omega)$ for relating the excitation $\bm{X}(\omega)$ to the modal coordinates $\bm{Q}(\omega)$,
\begin{align} \label{eq:F_SD_RelatingXandQ}
	\bm{Q}(\omega) = \underbrace{\overbrace{\left( -\omega^2 \bm{m} + i \omega \bm{d} + \bm{k}\right)^{-1}}^{\bm{H}(\omega)} \bm{\Phi}^{(x)}}_{H^{\bm{(xq)}}(\omega)} \bm{X}(\omega)
\end{align}
where the diagonal elements of the diagonal matrix $\bm{H}(\omega)$ represent the response functions of the SDOF system of each individual mode:
\begin{align} \label{eq:H}
	\bm{H}(\omega) = \left( \begin{array}{ccc} \ddots\\ &m_r(\omega_r^2+2\,i\,\xi_r\,\omega_r\,\omega-\omega^2)&\\ &&\ddots\\ \end{array}\right)
\end{align}
Relevant for fatigue assessments are modal stresses $\bm{\Phi}^{(\bm{\sigma})}$ and strains $\bm{\Phi}^{(\bm{\varepsilon})}$  --- the mode shapes of stress resp.~strain. These can be derived by applying the differential operator matrix $\bm{S}$ and the elastic matrix $\bm{E}$ \cite{Kranjc.2016}, so that for $r$-th mode ($r = 1,...,N$)
\begin{subequations}\label{eq:F_SD_ModesShapes}
	\begin{align} 
		\bm{\phi}_r^{(\bm{\varepsilon})} &= \bm{S} \bm{\phi}_r  \label{subeq:F_SD_ModesShapesStrain} \\
		\bm{\phi}_r^{(\bm{\sigma})} &= \bm{E} \bm{\phi}_r^{(\bm{\varepsilon})}
		\label{subeq:F_SD_ModesShapesSigma}
	\end{align}
\end{subequations}
Consequently, stress responses can be obtained by modal superposition, whereby $\bm{\Phi}^{(\bm{\sigma})}$ aggregates the relevant modal stresses (Eq.~\ref{subeq:F_SD_ModesShapesSigma})
\begin{align} \label{eq:F_SD_StressSolutionModelSuperposition}
	\bm{\sigma}(t) = \bm{\Phi}^{(\bm{\sigma})} \bm{q}(t) = \sum_{r}^{N} \bm{\phi}_r^{(\bm{\sigma})} q_r(t)
\end{align}
In the same way, using modal superposition, we can derive matrices of frequency response functions (FRFs), e.g.~relating the excitation $\bm{x}(t)$ to the stress response $\bm{\sigma}(t)$ by $H^{(\bm{x}\bm{\sigma})}(\omega)$. These FRF matrices have the dimension $(U = N_x) \text{ x } (V = N_\sigma)$, relating the number of loads $N_x$ with the number of stress components $N_\sigma$; $N_\sigma = 6$ for 3D resp.~$N_\sigma=3$ for 2D elements,
\begin{align} \label{eq:F_SD_FRFofStress}
	H^{(\bm{x}\bm{\sigma})}(\omega) = \bm{\Phi}^{(\bm{\sigma}) } \bm{H}(\omega) \bm{\Phi}^{(\bm{x})}
\end{align}
Alternatively, making use of Eq.~\eqref{eq:F_SD_RelatingXandQ}, \cite{Braccesi.2016} proposed to accelerate the computation of response PSDs by
\begin{subequations}\label{eq:MA_StressPSDViaBraccesi}
	\begin{align} 
		G_2^{(\bm{\sigma})}(\omega) &= \bm{\Phi}^{(\bm{\sigma})^T} G_2^{(\bm{q})}(\omega) \bm{\Phi}^{(\bm{\sigma})} \label{subeq:MA_StressPSDViaBraccesi1} \\
		G_2^{(\bm{q})}(\omega) &= H^{(\bm{x}\bm{q})^T}(\omega) G_2^{(\bm{x})}(\omega) H^{(\bm{x}\bm{q})}(\omega) \label{subeq:MA_StressPSDViaBraccesi2}
	\end{align}
\end{subequations}
resp.~in the subsequent paper \cite{Braccesi.2017} to modal spectral moments, relevant for a statistical-based fatigue assessment using PSD-based damage and load spectrum estimators. The central advantage lies in that between the PSD of modal coordinates $G_2^{(\bm{q})}(\omega)$ and the PSD of stresses $G_2^{(\bm{\sigma})}(\omega)$ stands only a matrix multiplication but no more frequency-dependent characteristic. \\
To conclude the fundamentals section and set the scope of the main section --- we can define different paths to obtain response kurtoses $\beta^{(\bm{y})}$ resp.~$\beta^{(\bm{\sigma})}$, giving insight into the impact of non-Gaussian loading on linear time-invariant systems. The most straightforward path towards response kurtoses is by acquiring the response time series (Fig.~\ref{fig:Approaches4DerivingReponseKurtoses}c,d), either using FRFs (e.g.~on the basis of modal analysis, Eq.~\ref{eq:F_SD_FRFofStress}) or by directly scaling the modal solution by its related mode shapes Eq.~\eqref{eq:F_SD_StressSolutionModelSuperposition}. Subsequently, the relevant moments can be estimated in the time-domain (Eqs.~\ref{eq:F_UV_NOrderCentMoments}, \ref{eq:F_UV_Beta}). Other possibilities are estimating the corresponding spectral densities (Fig.~\ref{fig:Approaches4DerivingReponseKurtoses} (a,b); Eq.~\ref{eq:F_UV_Kurtosis}), which after estimation, can be transferred for the corresponding response spectra \cite{Trapp.2021b,Trapp.2021g} in the frequency domain. The scalar-valued moments are calculated by integration, resp.~numerically by summation (Eq.~\ref{eq:F_UV_HOParsvl}). Note that in this approach, not only the scalar-valued moments are obtained, but also their spectral decompositions --- providing critical information by resolving the moments for frequency. The following main section presents the remaining and most efficient approach (Fig.~\ref{fig:Approaches4DerivingReponseKurtoses}e, red).
\section{Modal approach for fourth-order statistical characterizations}
\label{sec:Main_MA4Kurtosis}
In this section, we extend the multivariate analysis to moments and cumulants of fourth-order (Sec.~\ref{subsec:4M_MomsAndCums}), before this statistical characterization is linked to modal analysis in structural dynamics (Sec.~\ref{subsec:4M_ModalApproach}).
\subsection{Multivariate fourth-order moment and cumulant}
\label{subsec:4M_MomsAndCums}
Extending the multivariate analysis to fourth-order bears the challenge of higher-order $n>2$ multi-dimensionality. Nevertheless, this is necessary for characterizing modal solutions, which is the prerequisite for obtaining response characterizations, such as fourth-order responses. Naturally, this fourth-order multivariate characterization not only applies to the modal solution, but to all correlated multivariate processes, e.g.~in- $\bm{x}$ and outputs $\bm{y}$ of a dynamic system. While the multivariate second-order moment --- the covariance matrix --- is a tensor of rank two, the multivariate fourth-order moment extends Eq.~\eqref{eq:F_MVM_CovMatrix_Tensor} to a tensor of rank four that has $U^4$ elements
\begin{align} \label{eq:4M_MaC_4thMomentTensor}
\begin{split}
\mu_4^{(\bm{x})} = E\big[(\bm{X}(t)-E[\bm{X}(t)])^{4}\big] = E\big[(\bm{x}-\mu^{(\bm{x})})^{\otimes 4}\big] \\
= E\big[(\bm{x}-\mu^{(\bm{x})}) \otimes (\bm{x}-\mu^{(\bm{x})}) \otimes (\bm{x}-\mu^{(\bm{x})}) \otimes (\bm{x}-\mu^{(\bm{x})}) \big]
\end{split}
\end{align}
and four subscripts $u_1,...,u_4$ accordingly. This tensor of rank four is cumbersome to visualize and also contains a large set of symmetries. Therefore, we map $\mu_4^{(\bm{x})}$ to a tensor of rank two with the help of the Voigt form, that can represent symmetric tensors by reducing its rank. In in a first step, Eq.~\eqref{eq:4M_MaC_CovVoigt} expresses the covariance matrix $\bm{\Sigma}$ in Voigt form $\tilde{\bm{\Sigma}}$, whereby we extend the vector form following the rows of the upper triangle (this might deviate from other definitions, where the sequence of the off-diagonal elements is flipped) so that: 
\begin{equation} \label{eq:4M_MaC_CovVoigt}
\tilde{\bm{\Sigma}}^{(\bm{x})} = \big[\underbrace{\mu_2^{(x_1)},...,\mu_2^{(x_U)}}_{U \text{ elements}},\underbrace{\mu_2^{(x_{1,2})},...,\mu_2^{(x_{1,U})},\mu_2^{(x_{2,3})},...,...,\mu_2^{(x_{U-1,U})}}_{{Q=((U-1)U)/2 \text{ elements}}}\big]
\end{equation}  
In this process the symmetry of the covariance matrix is eliminated, which in Eq.~\eqref{eq:4M_MaC_CovVoigt} manifests in that cross entries $x_{u_1,u_2}$ appear in the form of $u_1 < u_2$, so that $\tilde{\bm{\Sigma}}$ has $(U+Q) = \frac{(U+1)U}{2}$ unique elements. By extending this to fourth-order gives the projected 2D-representation of the rank four tensor $\hat{\mu}_4^{(\bm{x})}$ that has $(U+Q)$ x $(U+Q)$ resp.~$\frac{(U+1)U}{2}$ x $\frac{(U+1)U}{2}$ elements (comp.~\cite{Basser.2007,Moakher.2008,Hine.2009})
\begin{equation} \label{eq:4M_MaC_Mom4ByCovAnalysis}
\begin{split}
\hat{\mu}_4^{(\bm{x})} = \begin{pmatrix}
\begin{array}{c|c}
U \text{ x } U  & U \text{ x } Q \\
\text{(co-)variances of variances}  & \text{covariances of co- and variances} \\
\begin{bmatrix}
\mu_{4}^{(x_1x_1)} & \cdots & \mu_{4}^{(x_1x_U)} \\
\vdots  & \ddots & \vdots \\
\mu_{4}^{(x_Ux_1)} & \cdots & \mu_{4}^{(x_Ux_U)}  \\
\end{bmatrix}
&
\begin{bmatrix}
 \mu_{4}^{(x_1x_{1,2})} & \cdots  & \mu_{4}^{(x_1x_{U-1,U})} \\
\vdots  & \ddots & \vdots \\
\mu_{4}^{(x_Ux_{1,2})} & \cdots & \mu_{4}^{(x_Ux_{U-1,U})} \\
\end{bmatrix} \\
\midrule
Q \text{ x } U  & Q \text{ x } Q \\
\text{covariances of co- and variances}  & \text{(co-)variances of covariances}  \\
\begin{bmatrix}
\mu_{4}^{(x_{1,2}x_1)} & \cdots & \mu_{4}^{(x_{1,2}x_U)} \\
\vdots  & \ddots & \vdots \\
\mu_{4}^{(x_{U-1,U}x_1)} & \cdots & \mu_{4}^{(x_{U-1,U}x_U)} \\
\end{bmatrix}
&
\begin{bmatrix}
\mu_{4}^{(x_{1,2}x_{1,2})} & \cdots & \mu_{4}^{(x_{1,2}x_{U-1,U})} \\
\vdots  & \ddots & \vdots \\
\mu_{4}^{(x_{U-1,U}x_{1,2})} & \cdots & \mu_{4}^{(x_{U-1,U}x_{U-1,U})} \\
\end{bmatrix}
\end{array}
\end{pmatrix} 
\end{split}
\end{equation}
Eq.~\eqref{eq:4M_MaC_Mom4ByCovAnalysis} defines four sub-matrices that lend to different interpretation in respect to the covariance matrix (Eq.~\ref{eq:F_MVM_CovMatrix}). Therefore, we use the notation $x_{u_1,u_2}x_{u_3,u_4} \hat{=} x_{u_1}x_{u_2}x_{u_3}x_{u_4}$, grouping the first and the second pair of subscripts.
When comparing sizes of the second- and fourth-order moment matrices, $\mu_4^{(\bm{x})}$ grew to $(U+Q) \text{ x } (U+Q)$ in comparison to the second-order covariance matrix ($U \text{ x } U$). However, we analogously find entries with same subscripts on the diagonal and different indices on the off-diagonals. \\
For defining a kurtosis tensor --- the multivariate extension of Eq.~\eqref{eq:F_UV_Beta} --- we need a denominator that represents the fourth-order moment of a multivariate stationary Gaussian process. Its representation corresponds to Eq.~\eqref{eq:4M_MaC_Mom4ByCovAnalysis} and is defined by 
\begin{equation} \label{eq:4M_MaC_Mom4Stat}
\hat{\mu}_{4,\text{stat}}^{(\bm{x})} = 
\begin{bmatrix}
\mu_{4,\text{stat}}^{(x_1x_1)} & \cdots & \mu_{4,\text{stat}}^{(x_1x_U)} & \mu_{4,\text{stat}}^{(x_1x_{1,2})} & \cdots  & \mu_{4,\text{stat}}^{(x_1x_{U-1,U})} \\
\vdots  & \ddots & \vdots & \vdots  & \ddots & \vdots \\
\mu_{4,\text{stat}}^{(x_Ux_1)} & \cdots & \mu_{4,\text{stat}}^{(x_Ux_U)} & \mu_{4,\text{stat}}^{(x_Ux_{1,2})} & \cdots & \mu_{4,\text{stat}}^{(x_Ux_{U-1,U})} \\
\mu_{4,\text{stat}}^{(x_{1,2}x_1)} & \cdots & \mu_{4,\text{stat}}^{(x_{1,2}x_U)} & \mu_{4,\text{stat}}^{(x_{1,2}x_{1,2})} & \cdots & \mu_{4,\text{stat}}^{(x_{1,2}x_{U-1,U})} \\
\vdots  & \ddots & \vdots & \vdots  & \ddots & \vdots \\
\mu_{4,\text{stat}}^{(x_{U-1,U}x_1)} & \cdots & \mu_{4,\text{stat}}^{(x_{U-1,U}x_U)} & \mu_{4,\text{stat}}^{(x_{U-1,U}x_{1,2})} & \cdots & \mu_{4,\text{stat}}^{(x_{U-1,U}x_{U-1,U})} \\
\end{bmatrix} \\
\end{equation}
where
\begin{equation} \label{eq:4M_MaC_Mom4StatIndividual}
\begin{split}
\mu_{4,\text{stat}}^{(x_{u_1,u_2}x_{u_3,u_4})} =
\mu_{2}^{(x_{u_1,u_2})} \mu_{2}^{(x_{u_3,u_4})} + \mu_{2}^{(x_{u_1,u_3})} \mu_{2}^{(x_{u_2,u_4})} + \mu_{2}^{(x_{u_1,u_4})} \mu_{2}^{(x_{u_2,u_3})} 
\end{split}
\end{equation}
three terms contribute that relate to combinatorial combinations of co- and variances, commonly referred to as Isserlis theorem \cite{Isserlis.1918}. While the fourth-order multivariate cumulant can be defined as their difference, 
\begin{equation} \label{eq:4M_MaC_Cum4}
c_{4}^{(\bm{x})} = \mu_{4}^{(\bm{x})} - \mu_{4,\text{stat}}^{(\bm{x})}; \hspace{10mm} \hat{c}_{4}^{(\bm{x})} = \hat{\mu}_{4}^{(\bm{x})} - \hat{\mu}_{4,\text{stat}}^{(\bm{x})}; 
\end{equation}
we can now also designate a kurtosis tensor,
\begin{equation} \label{eq:4M_MaC_KurtosisTensor}
\beta^{(\bm{\sigma})} = \mu_{4}^{(\bm{\sigma})} \oslash \mu_{4,\text{stat}}^{(\bm{\sigma})}; \hspace{10mm} \hat{\beta}^{(\bm{\sigma})} = \hat{\mu}_{4}^{(\bm{\sigma})} \oslash \hat{\mu}_{4,\text{stat}}^{(\bm{\sigma})}    
\end{equation}
where $\oslash$ implies elementwise division. \\
To conclude the introduction of the fourth-order moment tensor --- note, that a consequent projection from $\mu_4^{(\bm{x})}$ to $\hat{\mu}_4^{(\bm{x})}$ (Eqs.~\ref{eq:4M_MaC_4thMomentTensor}, \ref{eq:4M_MaC_Mom4ByCovAnalysis}), must account for the symmetric elements that were eliminated because of this projection. To make this a valid tensor projection, in that it complies with the invariants of rank four and affine transformations, Eq.~\eqref{eq:4M_MaC_Mom4ByCovAnalysis} must be complemented by prefactors $\{ \sqrt{2},2\}$ 
\begin{equation} \label{eq:4M_MaC_Mom4Norm}
\hat{\mu}_{4,\text{norm}}^{(\bm{x})} = \begin{pmatrix}
\begin{array}{c|c}
\begin{bmatrix}
\mu_{4}^{(x_1x_1)} & \cdots & \mu_{4}^{(x_1x_U)} \\
\vdots  & \ddots & \vdots \\
\mu_{4}^{(x_Ux_1)} & \cdots & \mu_{4}^{(x_Ux_U)}  \\
\end{bmatrix}
&
\sqrt{2}\begin{bmatrix}
 \mu_{4}^{(x_1x_{1,2})} & \cdots  & \mu_{4}^{(x_1x_{U-1,U})} \\
\vdots  & \ddots & \vdots \\
\mu_{4}^{(x_Ux_{1,2})} & \vdots & \mu_{4}^{(x_Ux_{U-1,U})} \\
\end{bmatrix} \\
\midrule
\sqrt{2}\begin{bmatrix}
\mu_{4}^{(x_{1,2}x_1)} & \cdots & \mu_{4}^{(x_{1,2}x_U)} \\
\vdots  & \ddots & \vdots \\
\mu_{4}^{(x_{U-1,U}x_1)} & \cdots & \mu_{4}^{(x_{U-1,U}x_U)} \\
\end{bmatrix}
&
2\begin{bmatrix}
\mu_{4}^{(x_{1,2}x_{1,2})} & \cdots & \mu_{4}^{(x_{1,2}x_{U-1,U})} \\
\vdots  & \ddots & \vdots \\
\mu_{4}^{(x_{U-1,U}x_{1,2})} & \cdots & \mu_{4}^{(x_{U-1,U}x_{U-1,U})} \\
\end{bmatrix}
\end{array}
\end{pmatrix} 
\end{equation}
\subsection{Modal approach for statistical response analysis} \label{subsec:4M_ModalApproach}
In this section the fourth-order multivariate analysis is coupled with the modal approach. It presents how the multivariate fourth-order moment and subsequently the kurtosis can be derived efficiently for full FE models with minimal computational effort. \\
In practical applications, it has been established that fatigue evaluations are based on Eq.~\eqref{eq:F_SD_FRFofStress}, using the FRF matrices that relate the dynamic input loads to the stress responses. These are subsequently used to carry out response analysis individually for each node or element of interest looping through the FE model, which then is followed by the subsequent post-processing steps of a fatigue assessment. This bears significant computational effort. However, Equation \eqref{eq:F_SD_StressSolutionModelSuperposition} hints at a more efficient way, which is to calculate the modal solution $\bm{q}(t)$, compromising a single response analysis, which is then used to obtain all other physical quantities. Therefore, we propose a general approach for obtaining arbitrary statistical response characterizations in a highly efficient way, following the steps of (\textit{i}) obtaining the modal solution, (\textit{ii}) estimating the desired statistical characterizations of the modal solution and (\textit{iii}) scaling these by the mode shapes for fast response analyses (Figure \ref{fig:Approaches4DerivingReponseKurtoses}). Therefore, it requires the FRF matrix $H^{(\bm{x}\bm{q})}(\omega)$ relating the input loads $\bm{x}(t)$ with the modal coordinates $\bm{q}(t)$ (dimensions $N_x \text{ x } N_r$ over frequency $f$ resp.~angular frequency $\omega$, Eq.~\ref{eq:F_SD_RelatingXandQ}). The modal solution resulting from the input loads $\bm{x}(t)$ is obtained by, 
\begin{align} \label{eq:MA_ModalSolution}
\bm{q}(t) = \mathcal{F}^{-1} \{ \bm{Q}(f)\} =  \mathcal{F}^{-1} \{ \bm{H}^{(\bm{x}\bm{q})^T}(f) \bm{X}(f) \}
\end{align}
where $\mathcal{F}^{-1} \{\cdot\}$ is the inverse Fourier transform, providing the particular modal solution. Now, Eq.~\eqref{eq:F_SD_StressSolutionModelSuperposition} can be used to gather the time series of stresses $\bm{\sigma}(t)$ for each point of interest without any additional response analysis --- solely by scaling the modal solution $\bm{q}(t)$ with the relevant local mode shapes $\bm{\Phi}$ (Fig.~\ref{fig:Approaches4DerivingReponseKurtoses}e). \\
Furthermore, this can be extended to all statistical characterizations that can be estimated from the modal solution. Exemplary, we can estimate the covariance matrix of the modal solution $\bm{\Sigma}^{(\bm{q})} = \mu_2^{(\bm{q})}$ of size $N_r \text{ x } N_r$ and obtain the covariance matrix for each node resp.~element by
 \begin{align} \label{eq:MA_MA4Covariance}
\bm{\Sigma}^{(\bm{\sigma})} = \mu_2^{(\bm{\sigma})} =  \bm{\Phi}^{(\bm{\sigma})^T} \bm{\Sigma}^{(\bm{q})} \bm{\Phi}^{(\bm{\sigma})} = \bm{\Phi}^{(\bm{\sigma})^T}  \mu_2^{(\bm{q})} \bm{\Phi}^{(\bm{\sigma})}
\end{align}
It is critical, that these statistical characterizations encompass the full multivariate representation of $\bm{q}(t)$. Despite the modal decoupling, the off-diagonals entries of the modal covariance matrix $\bm{\Sigma}^{(\bm{q})}$ include central information that can significantly contribute to the result. A natural cause are time-dependent excitation of the individual modes stemming from spectral overlap. \\
Lastly and centrally, this can be extended to the fourth-order moment resp.~cumulant by
\begin{subequations} \label{eq:MA_MA4FourthOrderMoment}
\begin{align}
\mu_4^{(\bm{\sigma})} &=  \bm{\Phi}^{(\bm{\sigma})^T}\bm{\Phi}^{(\bm{\sigma})^T}  \mu_4^{(\bm{q})} \bm{\Phi}^{(\bm{\sigma})} \bm{\Phi}^{(\bm{\sigma})} \\
\mu_{4,\text{stat}}^{(\bm{\sigma})} &=  \bm{\Phi}^{(\bm{\sigma})^T}\bm{\Phi}^{(\bm{\sigma})^T}  \mu_{4,\text{stat}}^{(\bm{q})} \bm{\Phi}^{(\bm{\sigma})} \bm{\Phi}^{(\bm{\sigma})} \\
c_4^{(\bm{\sigma})} &=  \bm{\Phi}^{(\bm{\sigma})^T}\bm{\Phi}^{(\bm{\sigma})^T}  c_4^{(\bm{q})} \bm{\Phi}^{(\bm{\sigma})} \bm{\Phi}^{(\bm{\sigma})}
\end{align}
\end{subequations}
which transforms the $N_r^4$ dimensional moment and cumulant tensors into size $6^4$ (3D elements) resp.~$3^4$ (2D elements) tensors representing the stress responses statistically. Now this has substantial computational advantages over MIMO response analysis, whenever the number of relevant modes is limited to a reasonable number, i.e.~realistic FE analysis using a modal approach. The operations of Eq.~\eqref{eq:MA_MA4FourthOrderMoment} can be carried out making use of tensor algebra, which has been implemented in all standard programming language thanks to the rise of machine- and deep learning. This calculation (Eq.~\ref{eq:MA_MA4FourthOrderMoment}) is independent of the length of the input loads, solely depending on the number of modes $N_r$ considered and the number of nodes to be evaluated. Even for large models this is a matter of seconds with standard PCs. \\
Then, there are several options to proceed: (\textit{i}) For meaningfully representing and analyzing the response tensors, $\mu_4^{(\bm{\sigma})}$ can be projected into Voigt form according to Eq.~\eqref{eq:4M_MaC_Mom4ByCovAnalysis}; (\textit{ii}) The fourth-order moment of the stress series components $[\mu_4^{(\sigma_x)}, \mu_4^{(\sigma_y)}, \mu_4^{(\sigma_z)}, \mu_4^{(\sigma_{xy})}, \mu_4^{(\sigma_{xz})}, \mu_4^{(\sigma_{yz})}]$ can be obtained from the generalized diagonal of these tensors; (\textit{iii}) For analyzing the impact of the non-Gaussianity of loading, we can calculate the fourth-order cumulant $c_4^{(\bm{\sigma})}$ (Eq.~\ref{eq:4M_MaC_Cum4}) or the kurtosis tensors (Eq.~\ref{eq:4M_MaC_KurtosisTensor}); (\textit{iv}) And \cite{Trapp.2024b} presents how these tensors can be rotated to find critical planes of the stress tensors. \\ 
As a final remark, there are other options for applying the introduced theory. When confronted with non-Gaussian processes, we often base our analysis on recorded time series. As such, we presented a robust path, which foresees to obtain the realization of the modal solution before estimating the stochastic characterization. Naturally, we could also choose to estimate the statistical characterization of the dynamic systems inputs, the loads, and directly transfer these to the corresponding characterization of the modal solution (bound to a frequency-domain characterization, like the PSD or the trispectrum, Eq.~\ref{eq:F_SD_RelatingXandQ}). This would also be the natural choice, considering pre-defined PSDs (test standards) or harmonic loading. This corresponds to the proposal of \cite{Braccesi.2017} for obtaining the PSD (Eq.~\ref{subeq:MA_StressPSDViaBraccesi2}) resp.~spectral moments of the modal solution. In an application using time series, the authors of this paper have experienced the most robust results, i.e.~zero deviations between the different approaches (Fig.~\ref{fig:Approaches4DerivingReponseKurtoses}), when the proposed steps are followed, first obtaining the modal solution $\bm{q}(t)$ (realization), from which consequently the multivariate statistical characterization is estimated. For example when the PSD matrix is directly estimated from the modal solution $\bm{q}(t)$, in contrast to transferring it via Eqs.~\eqref{eq:F_SD_RelatingXandQ}, \eqref{eq:MA_StressPSDViaBraccesi}. Even though in most settings, the difference in outcomes between these different sequences of processing will not become significant. 
\section{Validation and visualization} \label{sec:Validation}
\begin{figure}[]
\centering
\includegraphics[keepaspectratio,width=\textwidth,height=\textheight]{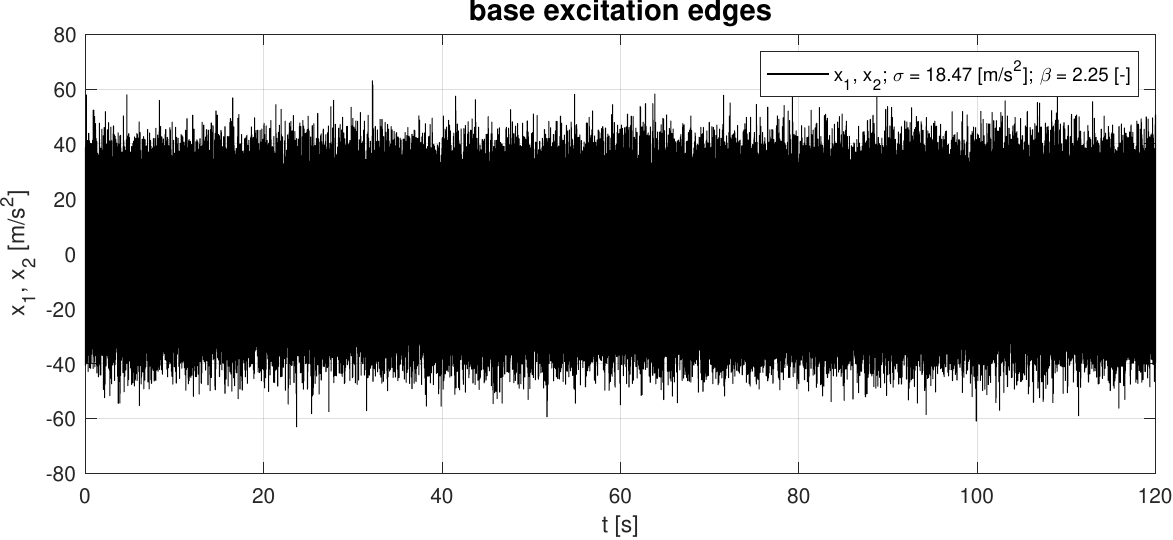}
 \caption{Sine-on-random load exciting the edges of the L-shaped specimen in $z$ direction}
\label{fig:Val_Excitation}
\end{figure}

\begin{figure}[]
\centering
\includegraphics[keepaspectratio,width=\textwidth,height=\textheight]{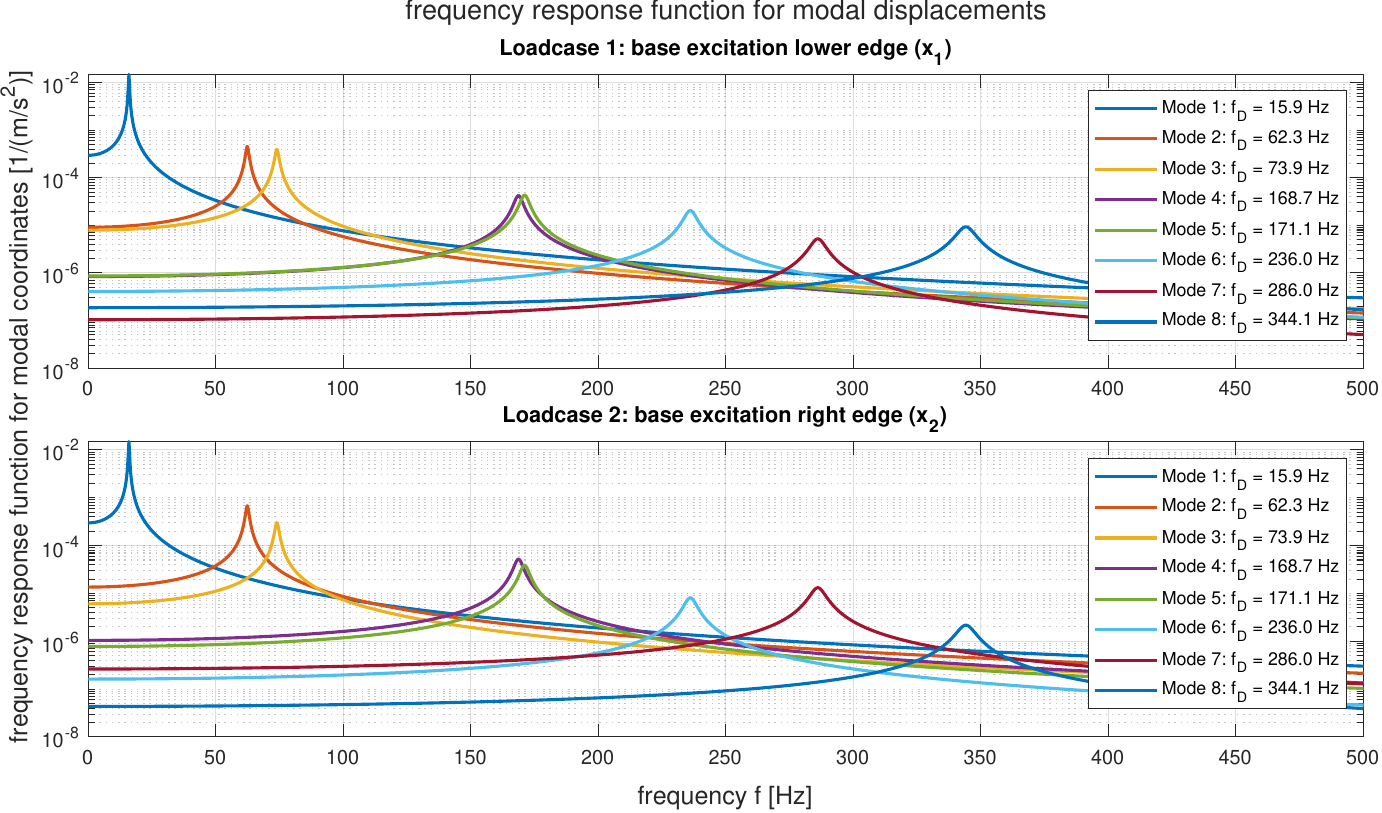}
 \caption{Transfer functions from clamped edges to modal solution $\bm{q}(t)$ with universal damping coefficient $\zeta = 0.01$}
\label{fig:Val_TF2ModalSolution}
\end{figure}

\begin{figure}[]
\centering
\includegraphics[keepaspectratio,width=\textwidth,height=\textheight]{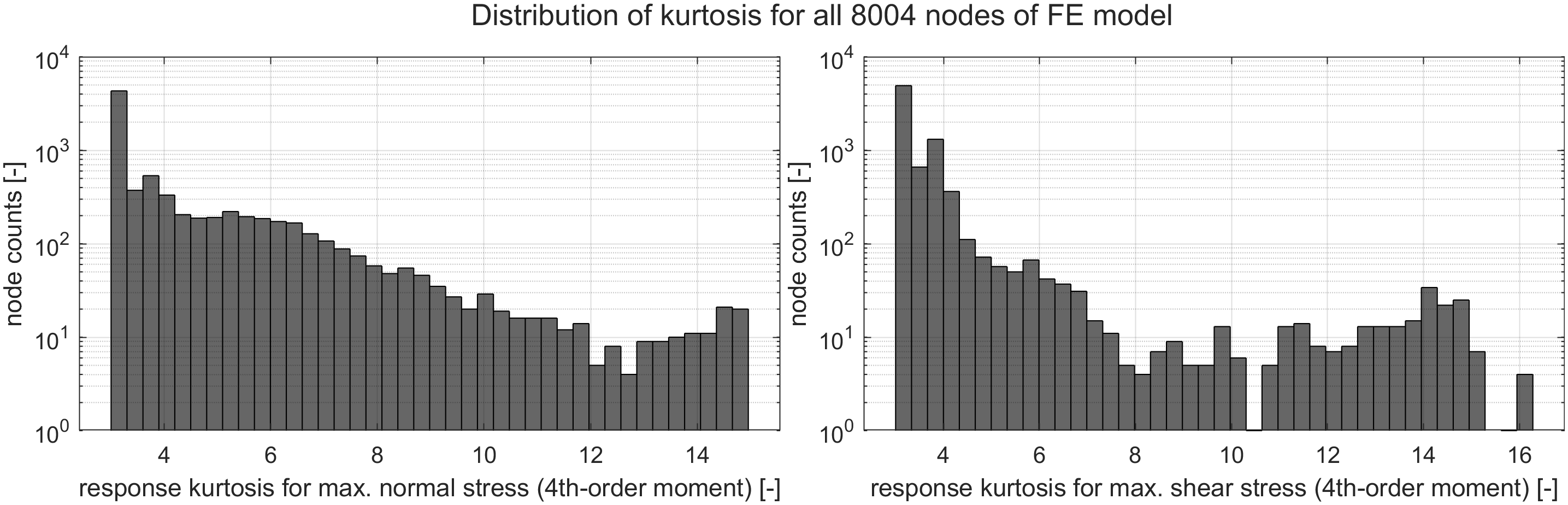}
 \caption{Distribution of kurtosis over nodes for maximum of stress components}
\label{fig:Val_Kurt4NodesLog}
\end{figure}

\begin{figure}[]
\centering
\includegraphics[keepaspectratio,width=0.75\textwidth,height=\textheight]{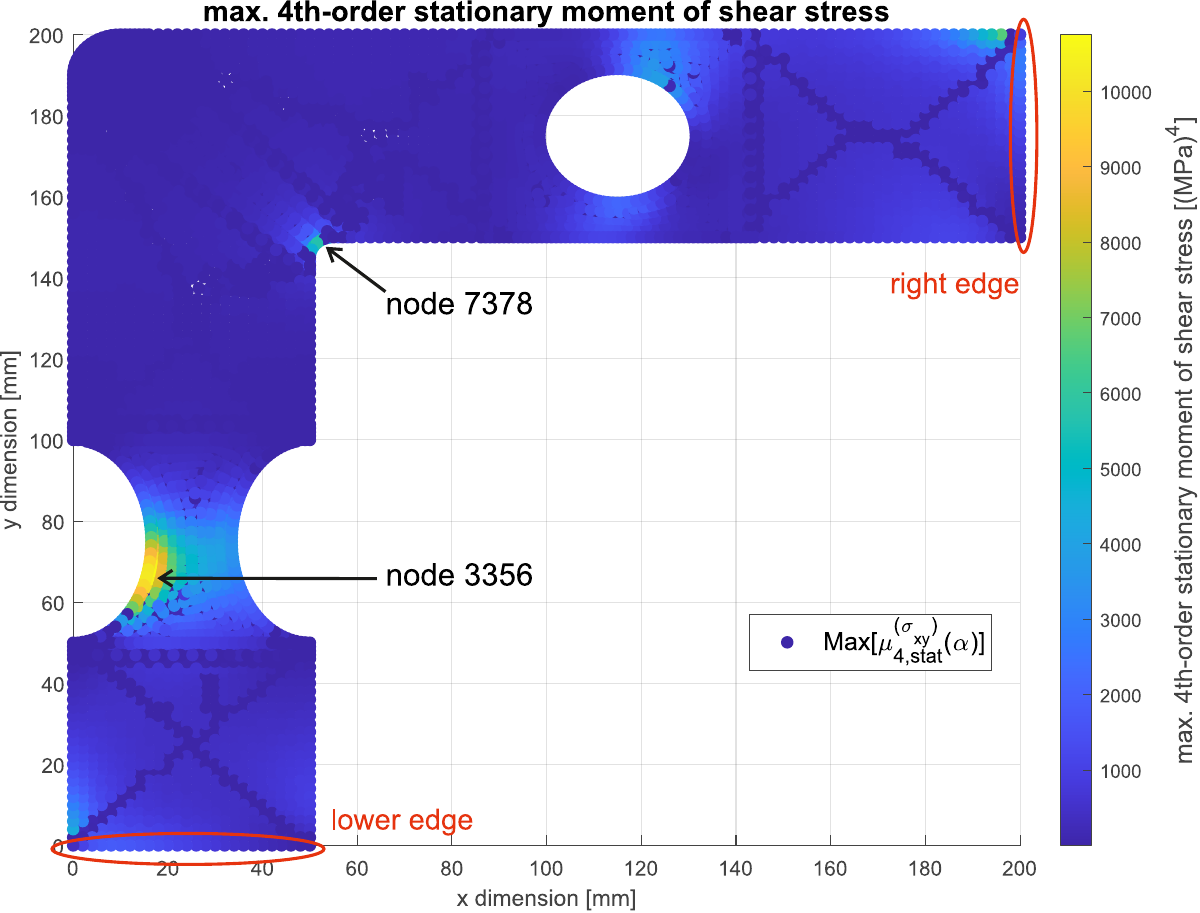}
 \caption{Maximum stationary fourth-order moment of shear stress (proportional to variance)}
\label{fig:Val_L_Mom4Stat}
\end{figure}

\begin{figure}[]
\centering
\includegraphics[keepaspectratio,width=0.75\textwidth,height=\textheight]{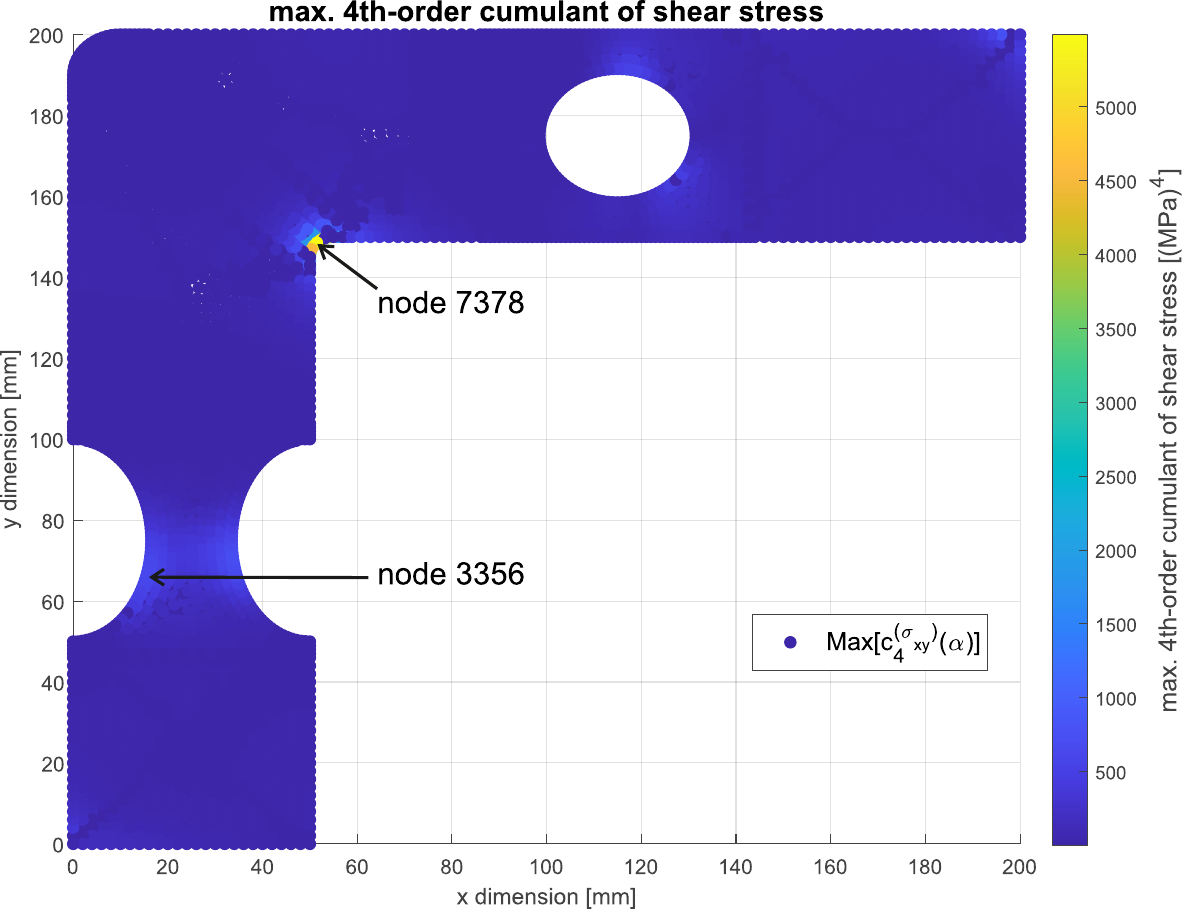}
 \caption{Maximum fourth-order cumulant of shear stress (pure non-Gaussian contribution to shear stress)}
\label{fig:Val_L_C4}
\end{figure}

\begin{figure}[]
\centering
\includegraphics[keepaspectratio,width=0.75\textwidth,height=\textheight]{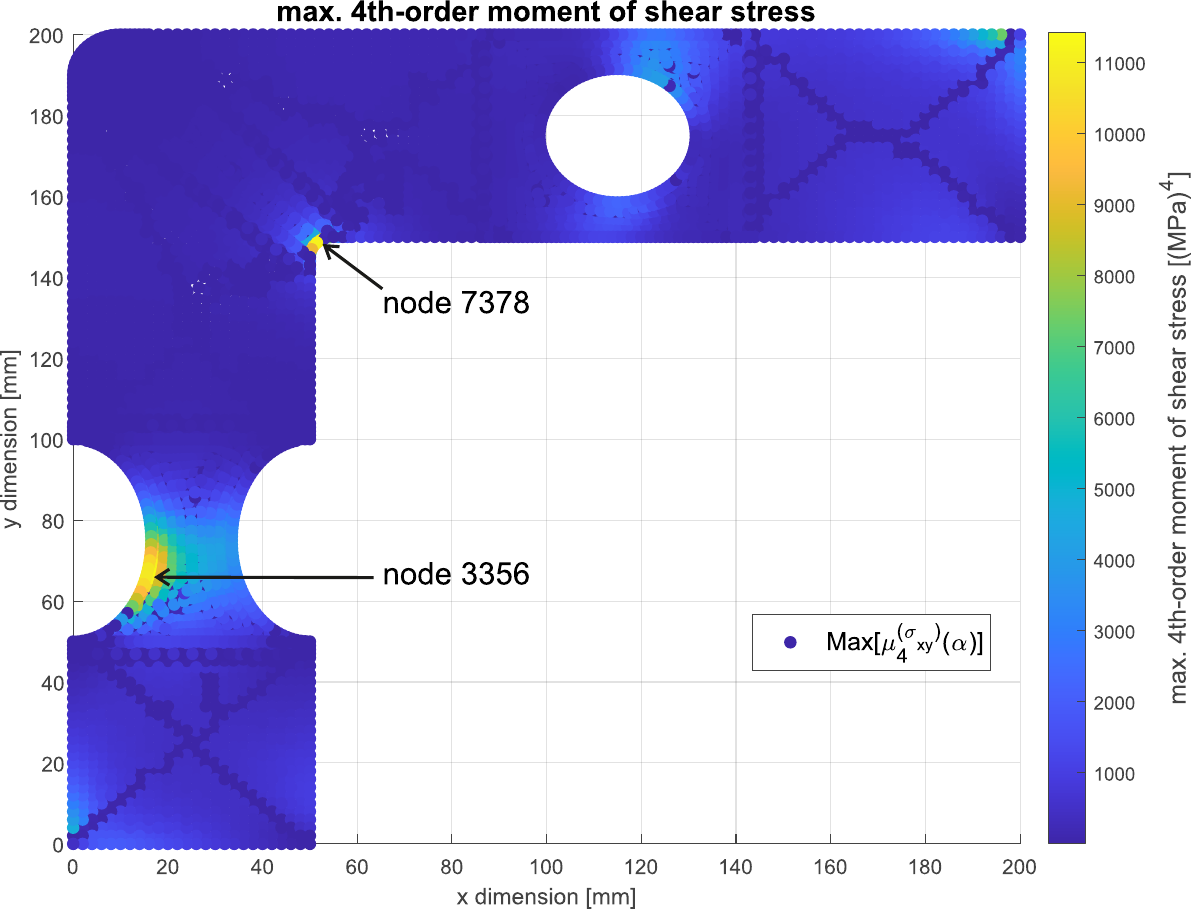}
 \caption{Maximum fourth-order moment of shear stress}
\label{fig:Val_L_Mom4}
\end{figure}

\begin{figure}[]
\centering
\includegraphics[keepaspectratio,width=\textwidth,height=\textheight]{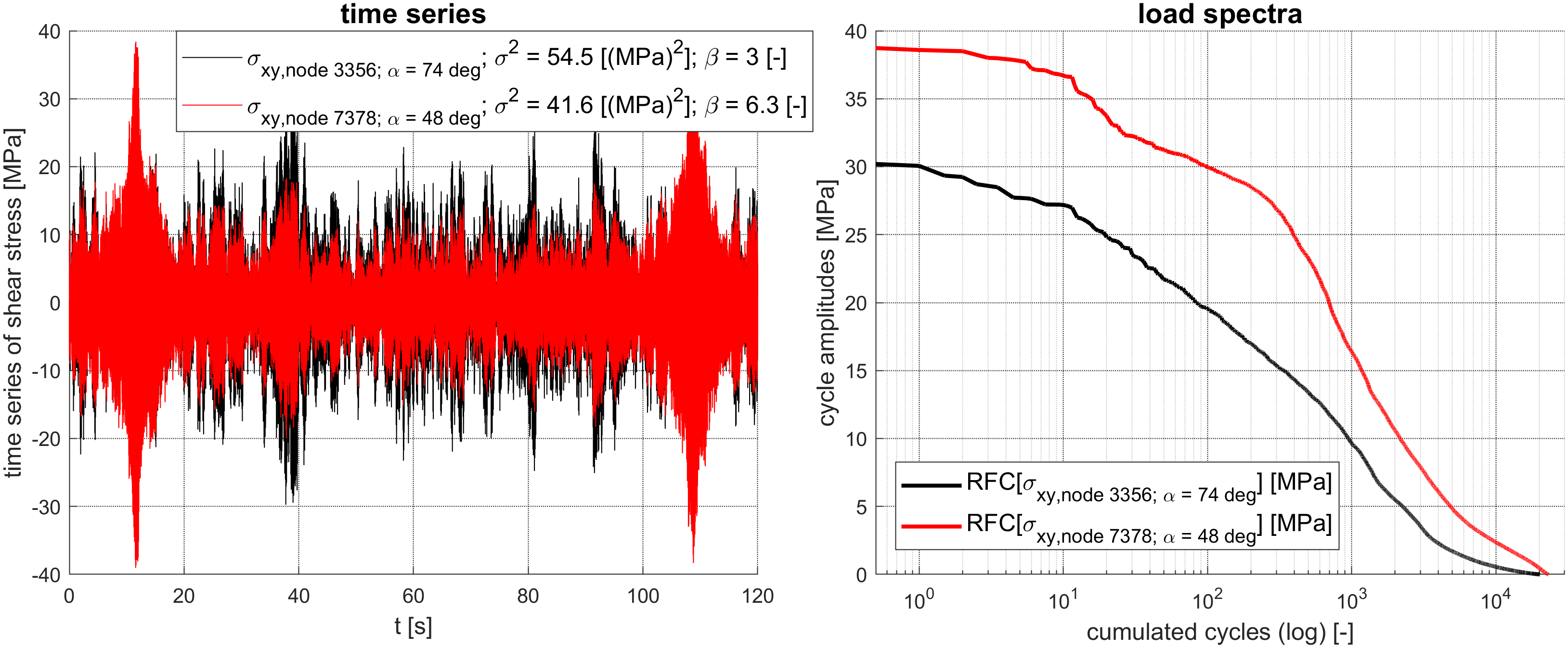}
 \caption{Comparison of critical nodes for shear stress by their time series of stress and load spectra}
\label{fig:VAL_L_responseSeries}
\end{figure}

\begin{figure}[]
	\centering
	\includegraphics[keepaspectratio,width=\textwidth,height=\textheight]{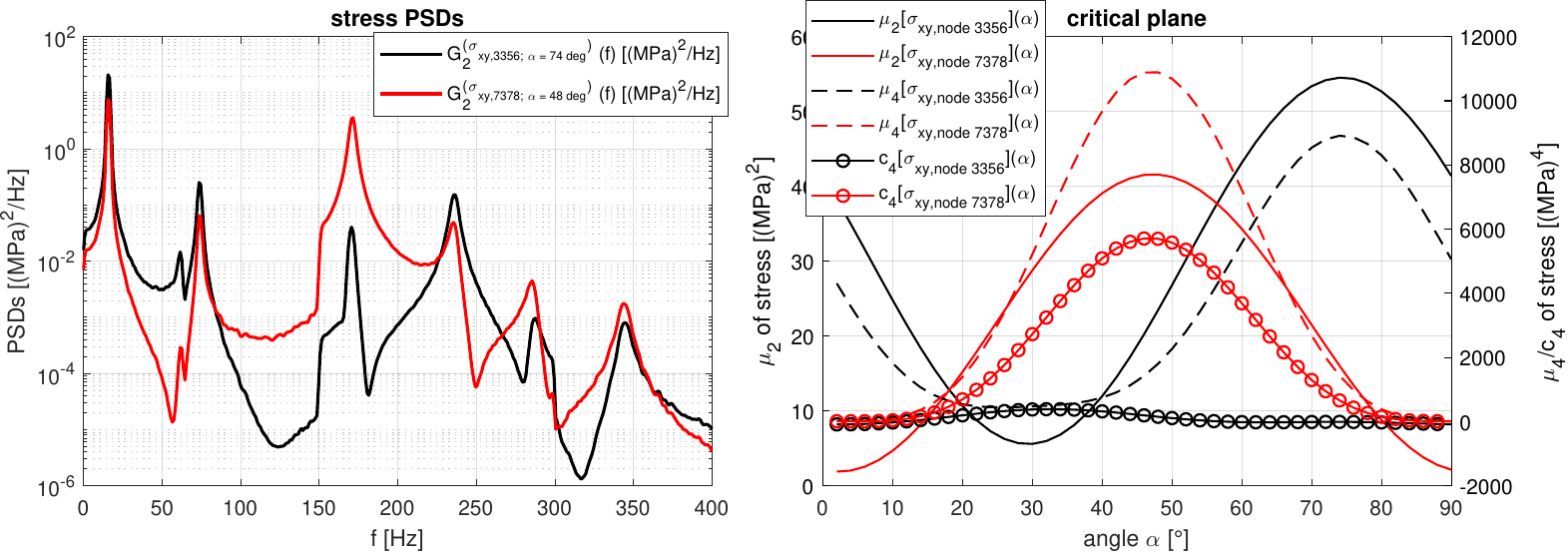}
	\caption{Comparison of critical nodes for their PSDs and critical orientation}
	\label{fig:VAL_L_responseSeriesPSDs}
\end{figure}

To provide exemplary results and to visualize the proposed approach, we apply the theory to a simple FE structure --- an 'L-shaped specimen' \cite{Pitoiset.2000} (Fig.~\ref{fig:Val_L_Mom4Stat}), which has found popular use for proposing and benchmarking methods in vibration fatigue. The L-shaped specimen has been intensively analyzed in \cite{Zhou.2019} including a comprehensive overview of its displacement and stress mode shapes. For the purpose of highlighting the characteristic of the approach presented in this paper, this structure is excited by a simple sine-on-random load, common for test standards across a wide set of technical fields. Some relevant test standards that apply such mixed-mode loading are defined for road vehicles (ISO 16750-3 \cite{InternationalOrganizationforStandardization.2023}; IEC 60068-2-80 \cite{InternationalElectrotechnicalCommission.2005}), military applications (MIL-STD-810 \cite{U.S.DepartmentofDefense.2008}), and airborne equipment (RTCA DO-160 \cite{RadioTechnicalCommissionforAeronautics.2010}). One of the central ideas of these mixed-mode load definitions is that in a variety of technical applications, vibration loads are composed of two essential components. On the one hand, sinusoidal vibration that may result from unbalanced mass forces in cylinders and rotors, forces that are transmitted through gears or pulse width modulation (PWM) of electric motors. On the other hand, all other vibration-schemes of machinery and environmental impacts that are represent by random noise with a given spectra density. Both, loading and structure, combine easy understanding and replicability, but nevertheless effectively underscore the effects of non-stationarity resp.~non-Gaussianity in the response of linear systems, which is the focus of this section. \\ 
Beginning with the loading, Figure \ref{fig:Val_Excitation} shows the time series sampled with rate $f_s = 2000$ Hz that is applied as base excitation to both clamped edges of the L-specimen (Fig.~\ref{fig:Val_L_Mom4Stat}) --- this set-up can be understood as resembling a classic shaker test. The load (Fig.~\ref{fig:Val_Excitation}) is composed of zero-mean random noise with a superimposed sine sweep. The random noise component has a standard deviation $\sigma = 10 \; 
[$m/s$^2]$ --- for a zero-mean process equivalent to its RMS level. The sine sweep has an amplitude of $A_{\text{sine}} = 22 \; [$m/s$^2]$ and sweeps between $[150, 300]$ Hz logarithmically with a rate of one octave/minute. The time series' kurtosis value of $\beta^{(x)} = 2.25$ (Fig.~\ref{fig:Val_Excitation}) lies exactly between the one of a pure sine wave ($\beta = 1.50$) and of random noise ($\beta = 3$). 
Figure \ref{fig:Val_TF2ModalSolution} shows the first eight modes of the 'L'-specimen that are considered in this analysis and depicts them by their individual FRF from each edge to the modal solution (Eq.~\ref{eq:F_SD_RelatingXandQ}). I.e.~while the excitation at the both edges are the same (Fig.~\ref{fig:Val_Excitation}), they have individual FRFs for the modal solution $\bm{q}(t)$. In this scenario, the modal solution is a multivariate process composed of $N_r = 8$ subprocesses --- the modal coordinates of each mode. After a single response analysis for obtaining the modal solution, all subsequent evaluations are gathered by mode shape scaling and modal superposition. We make use of this computational efficient approach for the following comprehensive analysis: Calculating the covariance matrix $\bm{\Sigma}^{(\bm{\sigma})}$ (Eq.~\ref{eq:MA_MA4Covariance}), the fourth-order moment $\mu_4^{(\bm{\sigma})}$, the stationary fourth-order moment $\mu_{4,\text{stat}}^{(\bm{\sigma})}$, the fourth-order cumulant $c_4^{(\bm{\sigma})}$ (Eqs.~\ref{eq:MA_MA4FourthOrderMoment}), and the kurtosis tensor (Eq.~\ref{eq:4M_MaC_KurtosisTensor}) for each single node of the FE model. These characterizations serve in this section for the analysis of structural response behavior as well as for the identification of critical nodes. By default, these are given in global coordinates, corresponding to the mode shape definition of the FE solver. Therefore, in a second step all nodal tensors are rotated using an increment of $\Delta \alpha = 2^\circ$ to find the angle of maximum normal and shear stress \cite{Trapp.2024b} --- so that 90 tensors (180 divided by 2 degrees) are calculated and stored for each node (see also Fig.~\ref{fig:VAL_L_responseSeriesPSDs}). Making use of the herein presented approach that exploits the modal solution and the efficient implementation of tensor algebra in common programming languages, this comprehensive analysis is carried out in less than a second for the full set of 8004 nodes of the FE model using a standard portable PC. \\
Now, these results allow us to highlight the interplay between dynamic loading and structural dynamics --- Fig.~\ref{fig:Val_Kurt4NodesLog} shows the distribution of response kurtoses for all nodes in the projection of maximum normal resp.~shear stress. This makes apparent, that the loads' kurtosis of $\beta^{(x)} = 2.25$ lead to a variety of different response kurtoses ranging between $\beta^{(y)} = [2.9, 16.1]$. 
In a next step we take a closer look at these effects of non-Gaussianity resp.~non-stationarity on the FE model. For this discussion, we use the statistical characterization of the maximum shear stress. We depict the nodal solutions for $\text{Max}\,[\mu_{4,\text{stat}}^{(\sigma_{xy})}]$ (Fig.~\ref{fig:Val_L_Mom4Stat}), $\text{Max}\,[c_{4}^{(\sigma_{xy})}]$ (Fig.~\ref{fig:Val_L_C4}), and $\text{Max}\,[\mu_{4}^{(\sigma_{xy})}]$ (Fig.~\ref{fig:Val_L_Mom4}). These are all based on fourth-order, so they can directly be related with another. The stationary fourth-order moment $\text{Max}\,[\mu_{4,\text{stat}}^{(\sigma_{xy})}]$ is proportional to the squared second-order moment $\text{Max}\,[\mu_{2}^{(\sigma_{xy})}]$ resp.~the maximum value for shear in the covariance matrix (Eq.~\ref{eq:F_UV_Beta}). This means, Fig.~\ref{fig:Val_L_Mom4Stat} resembles the stationary Gaussian contribution to the fourth-order moment. This contour plot stands representative for the conventional PSD-based analysis, when its integral, the variance, is depicted. On the contrary, the fourth-order cumulant $c_4^{(\sigma_{xy})}$ displays the pure deviation from stationarity and Gaussianity. What becomes apparent from Fig.~\ref{fig:Val_L_C4} is that in comparison to where we find the maximum value Max[$\mu_{2}^{(\sigma_{xy})}$] resp.~in terms of fourth order $\text{Max}\,[\mu_{4,\text{stat}}^{(\sigma_{xy})}]$ (node 3356, Fig.~\ref{fig:Val_L_Mom4Stat}), its highest values focus on a different cross section --- the notch in the joint of the 'L' (node 7378). This can be explained by the non-stationary portion of the excitational load --- the sine sweep. While the lower frequent modes are connected to high modal stresses at the left notch, the higher frequent modes, which correspond to the sweep range belong to larger modal stress in the notch (comp.~\cite{Zhou.2019}). Consequently, the sine sweeps through the modes critical for the right angle notch at node 7378, while the random noise --- the stationary Gaussian portion --- excites all frequencies, of which the first mode (Fig.~\ref{fig:Val_TF2ModalSolution}; see also \ref{fig:VAL_L_responseSeriesPSDs}) responds with the highest stresses at the left notch (node 3356). To identify critical structural response behavior, we have to bring both contributions --- the Gaussian and the non-Gaussian --- together, considering the fourth-order moment. Fig.~\ref{fig:Val_L_Mom4} shows its distribution over all nodes, whereby the node with the highest value for shear stress is node 7378 in the joint of the 'L'-specimen. To examine this is more detail, we take a more detailed look at the two nodes determined by $\text{Max}\,[\mu_{4,\text{stat}}^{(\sigma_{xy})}]$ (node 3356) and $\text{Max}\,[\mu_{4}^{(\sigma_{xy})}]$ (node 7378), and show their time series in the projection of maximum shear stress (Fig.~\ref{fig:VAL_L_responseSeries}). This figure includes the time series, their kurtosis, and the load spectra. While node 7378 has a lower variance, its load spectrum is composed of higher and more amplitudes. The latter comes from the higher frequent spectral content (PSDs in Fig.~\ref{fig:VAL_L_responseSeriesPSDs}). Lastly, Figure \ref{fig:VAL_L_responseSeriesPSDs} shows on the right hand side for both nodal stresses the statistical characterizations over the angle of rotation. While node 3356 has the larger second-order moment of shear stress, it is barely affected by any non-Gaussian characteristic ($c_4$). In contrast, large portions of $mu_4$ for node 7378 stem from non-Gaussianity and surpass that of node 3356. To conclude, in the presented example, the more damaging node 7378 correlates with the highest fourth-order moment. 

\section{Conclusion}
This paper introduces a novel approach for rapidly evaluating the impact of non-Gaussian inputs on linear dynamic systems by leveraging the modal solution in response to a given (multivariate) input loading. Traditionally, assessing the effects of non-Gaussian inputs on linear systems through numerical simulations such as finite element (FE) models required conducting individual dynamic analyses for each element resp.~node. However, the proposed approach utilizes a single dynamic response analysis --- the modal solution --- from which individual responses can be efficiently derived by scaling with the mode shapes of interest. This process significantly reduces computational demands, allowing for the full evaluation of FE models with minimal effort. Importantly, both the conventional method when using modal superposition and our proposal yield identical results, with our approach providing a much faster alternative. \\
Fundamentally, in this paper we characterize the 'non-properties' of non-stationarity and non-Gaussianity --- the characteristics that statistical response analyses so far not cover, using the fourth-order moment and its standardized representation, the kurtosis. For their highly efficient implementation in response analysis using FE models, we introduce the multivariate fourth-order moment and kurtosis tensor, which is the prerequisite for characterizing a dynamic structure's modal solution under arbitrary dynamic loading. Subsequently the fourth-order moment tensor at each element and node can be obtained by scaling the modal solution by its relevant mode shapes. This allows for the fast assessment of all nodal displacements, strains, and stresses through mode shape scaling and modal superposition, reducing this analysis for full FE models to a matter of seconds. 

Our methodology is exemplified using a popular FE structure which has been commonly referenced in studies on structural dynamics and vibration fatigue. Through this example, we demonstrate how structural responses are influenced by the interplay between loading and structural dynamics. The fourth-order moment and its related cumulant and kurtosis, effectively account for non-Gaussian and non-stationary effects in the statistical characterization of loads and structural responses, such as stresses. By analyzing nodal stresses using the maximum fourth-order moment, we successfully identified the more critical node, highlighting the potential of our approach for quick simulations of long multivariate time series input, such as recorded loads, but also for quickly obtaining feedback loops when optimizing structural designs. Respectively, as represented by the presented example for efficiently simulating non-Gaussian definitions in test standards. 

The results of this paper pave the way for further research, some of which is already underway. In \cite{Trapp.2024c} we proposed a data-driven damage estimator for non-stationary loading and studied the relevance of the fourth-order moment's spectral content for fatigue analysis, recognizing that higher frequent spectral content indicates more damaging cycles derived from stress states (this effect can be seen in Figure \ref{fig:VAL_L_responseSeries}, where the higher frequent stress state has a 'fuller' load spectrum, i.e.~more cycles per amplitude). On the one hand, the herein proposed methodology provides the vehicle to employ this damage estimator and other related approaches in the most efficient manner. On the other hand, it suggests to extend this approach for the statistical characterization of the fourth-orders spectral content --- deriving non-stationarity matrices of responses using modal solutions \cite{Trapp.2024d}. In our opinion, this can also be the prerequisite for a load spectrum estimator for non-stationary stress states, which in contrast to a damage estimator allows to employ other linear damage accumulation rules than Palmgren-Miner elementary. Here this paper builds the basis for this forthcoming paper that proposes a load spectra estimator for non-stationary stress states \cite{D.Fraulin.2024}. Lastly, in \cite{Trapp.2024b} we develop transfer matrices for the multivariate fourth-order moment stress tensors, allowing to efficiently identify critical planes with maximum fourth-order moment.

\bibliographystyle{elsarticle-num} 
\bibliography{literature}

\newpage 
\end{document}